\documentclass[
reprint,
superscriptaddress,
amsmath,amssymb,
aps,pra
]{revtex4-2}

\usepackage[utf8]{inputenc}
\usepackage{graphicx}
\usepackage{dcolumn}
\usepackage{bm}
\usepackage{dsfont}
\usepackage{hyperref}
\usepackage[dvipsnames]{xcolor}
\usepackage{amsfonts,amsthm}
\usepackage{xfrac}
\usepackage[normalem]{ulem}

\renewcommand{\vec}[1]{\mathbf{#1}}
\newcommand{\di}{\mathrm{d}}

\usepackage{soul}

\usepackage{braket}
\graphicspath{{figures/}}

\renewcommand{\eqref}[1]{Eq.~(\ref{#1})}

\newcommand{\figref}[1]{Fig.~\ref{#1}}
\newcommand{\Figref}[1]{Figure~\ref{#1}}
\newcommand{\secref}[1]{section~\ref{#1}}

\begin{document}
	
\title{Cavity Quantum Electrodynamics with Atom Arrays in Free Space}

\author{David Castells-Graells} \email{david.castells@mpq.mpg.de}
\affiliation{Max-Planck-Institut f\"ur Quantenoptik, Hans-Kopfermann-Strasse 1, 85748 Garching, Germany.}
\affiliation{Munich Center for Quantum Science and Technology, Schellingstrasse 4, 80799 M\"unchen, Germany.}
\author{J. Ignacio Cirac}
\affiliation{Max-Planck-Institut f\"ur Quantenoptik, Hans-Kopfermann-Strasse 1, 85748 Garching, Germany.}
\affiliation{Munich Center for Quantum Science and Technology, Schellingstrasse 4, 80799 M\"unchen, Germany.}
\author{Dominik S. Wild}
\affiliation{Max-Planck-Institut f\"ur Quantenoptik, Hans-Kopfermann-Strasse 1, 85748 Garching, Germany.}
\affiliation{Munich Center for Quantum Science and Technology, Schellingstrasse 4, 80799 M\"unchen, Germany.}

\date{\today}

\begin{abstract}
    Cavity quantum electrodynamics (cavity QED) enables the control of light-matter interactions at the single-photon level, rendering it a key component of many quantum technologies. Its practical realization, however, is complex since it involves placing individual quantum emitters close to mirror surfaces within a high-finesse cavity. In this work, we propose a cavity QED architecture fully based on atoms trapped in free space. In particular, we show that a pair of two-dimensional, ordered arrays of atoms can be described by conventional cavity QED parameters. Such an atom-array cavity exhibits the same cooperativity as a conventional counterpart with matching mirror specifications even though the cavity coupling strength and decay rate are modified by the narrow bandwidth of the atoms. We estimate that an array cavity composed of $^{87}\mathrm{Rb}$ atoms in an optical lattice can reach a cooperativity of about $10$. This value can be increased suppressing atomic motion with larger trap depths and may exceed $10^4$ with an ideal placement of the atoms. To reduce the experimental complexity of our scheme, we propose a spatially dependent AC Stark shift as an alternative to curving the arrays, which may be of independent interest. In addition to presenting a promising platform for cavity QED, our work creates opportunities for exploring novel phenomena based on the intrinsic nonlinearity of atom arrays and the possibility to dynamically control them.
\end{abstract}

\maketitle

\section{Introduction}

Quantum networks have been a focus of research in the field of quantum optics as they hold great potential for applications in quantum communication~\cite{bouwmeester2000Physics,kimble2008Quantum}, quantum cryptography~\cite{pirandola2020Advances}, distributed quantum computing~\cite{copsey2003Scalable,duan2010Colloquium}, quantum metrology~\cite{komar2014quantum,khabiboulline2019optical}, and the study of exotic many-body systems~\cite{illuminati2006Light,noh2017Quantum}. Photons are ideal carriers of quantum information between nodes of a quantum network as they can travel over long distances while interacting weakly with the environment. A key challenge in realizing robust and scalable quantum networks thus lies in coherently controlling the interaction between matter and light at the single-photon level.
Due to the small optical cross-section of a dipole emitter and the diffraction limit, deterministic light-matter interaction is unfeasible with far-field optics in free space. Cavity quantum electrodynamics (cavity QED) overcomes these limitations
by placing the emitter inside an optical resonator~\cite{walther2006cavity,reiserer2015cavity}. Although high-fidelity, deterministic light-matter interaction has been demonstrated in a variety of cavity QED setups~\cite{ritter2012Elementary,thompson2013Coupling,tiecke2014Nanophotonica,welte2018PhotonMediateda}, incorporating these systems into large networks presents a major technological challenge owing to the complexity of placing individual emitters inside high-finesse optical resonators while maintaining the coherence of both components~\cite{chang2018Colloquium,menon2024Integrated}.

\begin{figure}[b!]
    \vspace{-10pt}
	\includegraphics[width=0.99\linewidth]{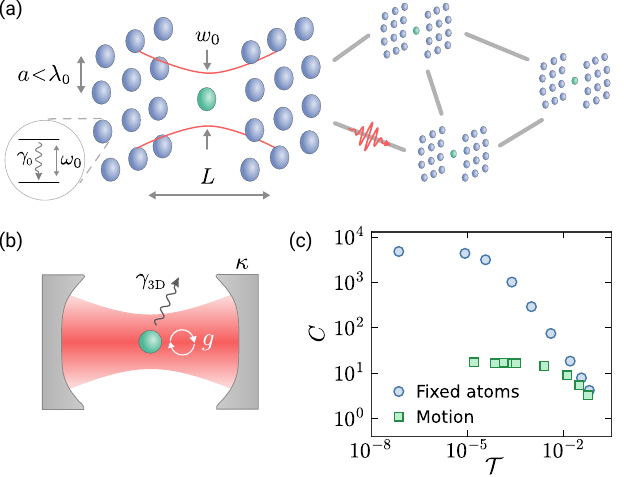}
    \vspace{-17pt}
	\caption{\label{figIntro} Cavity QED with atom arrays. (a)~An atom (green circle) interacts with the field between two subwavelength arrays (blue circles). The system efficiently absorbs and emits light, rendering it a suitable node of a quantum network. (b)~Atom-array cavities can be described in terms of conventional cavity QED parameters. (c)~Cooperativity of an atom-array cavity as a function of the mirror transmission, $\mathcal{T}$. The array mirrors, each composed of $30\times 30$ atoms, have lattice spacing $a=0.47\lambda_0$ and are curved to achieve a beam waist $w_0=2\lambda_0$. The mirror transmission is determined by the frequency of the cavity mode, which is tuned by varying the cavity length from $1.5 \lambda_0$ (smallest $\mathcal{T}$) to $1.54 \lambda_0$ (largest $\mathcal{T}$). We show the results for the ideal case of pinned atoms (blue circles) and including losses due to atomic motion (green squares) as computed in \secref{sec:motion} for $^{87}\mathrm{Rb}$ in an optical lattice with a depth of $2000$ recoil energies.
 }
\end{figure}

In this work, we propose an approach that addresses these shortcomings using the collective response of ordered arrays of emitters~\cite{asenjo2017exponential,reitz2022Cooperative,ruostekoski2023Cooperative}. For concreteness, we consider optical transitions in neutral atoms, although the approach readily applies to other types of dipole emitters. Previous works showed that a two-dimensional, ordered array of atoms with a subwavelength lattice spacing acts as a perfect reflector when the incident light is resonant with a collective excitation~\cite{bettles2016enhanced,shahmoon2017cooperative,rui2020subradiant}. Our proposal combines two such array mirrors to form an \emph{array cavity}, within which additional target atoms are placed as shown in \figref{figIntro}(a). The setup can be realized by trapping the atoms in an optical lattice far from any surface.

Similar setups without atoms inside the cavity were considered in previous studies, which identified narrow resonances in the reflection and transmission spectra due to long-lived collective excitations delocalized across the two arrays~\cite{guimond2019subradiant,pedersen2023quantum}. By studying the dynamics of an atom inside the array cavity, we show that these resonances can be viewed as discrete cavity modes. We find a large regime of parameters in which our setup is accurately described by a model of conventional cavity QED with a coupling strength $g$, cavity decay rate $\kappa$, and spontaneous emission rate $\gamma_\mathrm{3D}$ [see Fig.~\ref{figIntro}(b)]. Despite the similarities, the setup differs in important aspects from conventional cavity QED, arising from the narrow bandwidth over which the atom arrays reflect light. First, the array cavity only supports modes with frequencies near the resonance of the atoms that form it. Therefore, the array mirrors must be separated by a distance close to a half-integer multiple of their resonant wavelength. This is in contrast to a cavity formed by broadband mirrors, which supports modes at arbitrary separations of the mirrors. Secondly, the cavity QED parameters are modified as they are governed by the lifetime of the array atoms instead of the cavity round-trip time.

The modification of the cavity QED parameters can be understood from the fact that resonant reflection by a mirror with bandwidth $\Gamma_0$ incurs the Wigner time delay $\tau_\mathrm{delay} = 2 / \Gamma_0$~\cite{wigner1955lower,bourgain2013direct,pedersen2023nonlinear}. In the parameter regime considered here, $\tau_\mathrm{delay}$ far exceeds the propagation time $\tau_\mathrm{prop} = L/c$ of the photon traveling the distance $L$ between the arrays. The photon only spends a fraction $\tau_\mathrm{prop} / (\tau_\mathrm{prop} + \tau_\mathrm{delay}) \approx L \Gamma_0 / 2 c \ll 1$ of the time propagating and the number of round trips in a given time is reduced by the factor $\zeta = L \Gamma_0 / 2 c$ compared to broadband mirrors. It follows that $\kappa$ and $g^2$, the latter of which is proportional to the energy density of the electromagnetic field inside the cavity, are reduced by this factor. Intriguingly, both $g$ and $\kappa$ are therefore approximately independent of the cavity length $L$ in our setup. We highlight that the factors cancel when computing the cooperativity $C = 4 g^2 / \kappa \gamma_\mathrm{3D}$, which is a key figure of merit. An atom-array cavity thus achieves the same cooperativity as a conventional cavity QED setup with matching mirror specifications, i.e., the radius of curvature and the reflection and transmission coefficients.

While the atom-array cavity obviates the need for trapping near dielectric surfaces, it presents its own challenges related to the precise positioning of the atoms. As illustrated in Fig.~\ref{figIntro}(c), the cooperativity saturates when the transmission of the mirrors is less than the scattering loss. For atoms pinned at locations that match the curved wavefront of the cavity mode, the loss is small and cooperativities on the order of the $10^4$ can be achieved. Disorder and motion of the atoms significantly increase scattering loss. Nevertheless, cooperativities exceeding unity can be achieved with current experimental parameters. We show below that the requirement on the curved positioning of the atoms can be relaxed by subjecting a flat array to a spatially varying Stark shift.

The properties of atom arrays create new opportunities beyond existing cavity QED schemes. Atom array setups have been proposed to store light with high fidelity~\cite{facchinetti2016Storing,asenjo2017exponential,manzoni2018optimization,guimond2019subradiant,rubies-bigorda2022Photon}, modify the optical wavefront~\cite{ballantine2021Cooperative,fernandez-fernandez2022Tunable,bassler2024metasurface}, enhance absorption~\cite{higgins2014Superabsorption,holzinger2020Nanoscale,ballantine2022Unidirectional}, and mediate long-range interactions~\cite{masson2020Atomicwaveguidea,patti2021Controlling,castells2021atomic}. The combination of such applications with the setup presented here leads to a powerful toolbox for designing quantum networks with desirable attributes not accessible with conventional mirrors. The ability to optically trap atoms at relatively short distances can be used to explore more complex schemes within the same experimental setup, involving multiple atoms inside a cavity, or even multiple connected cavities. Another compelling feature is the ability to dynamically control the properties of atom arrays by means of external fields~\cite{weitenberg2011Singlespin,bloch2012Quantum,wang2015Coherent}. For instance, the mirrors could be rapidly switched on or off by optical control and the polarization of the optical transition could be modulated to create cavity modes with time-dependent chirality. Novel schemes could further take advantage of the motional degrees of freedom of the atoms within the framework of optomechanics~\cite{shahmoon2020Quantuma,shahmoon2020Cavity} or of the intrinsic nonlinearity of the arrays. Unlike standard nonlinear media~\cite{oshea2011Alloptical,carusotto2013Quantum}, atom arrays display nonlinearities at the level of few photons~\cite{mirhosseini2019Cavitya,bettles2020Quantum,masson2020Atomicwaveguidea} or even at the single-photon level in cavity configurations~\cite{pedersen2023quantum,robicheaux2023Intensity}. Finally, Rydberg states can be leveraged to realize photonic gates~\cite{bekenstein2020Quantum,moreno-cardoner2021Quantum,zhang2022Photonphoton,srakaew2023subwavelength}. Applications of the cavity setup involving Rydberg states might be of special interest, as they are highly susceptible to interfering fields from nearby surfaces~\cite{carter2012Electricfield}.

The paper is structured as follows. In \secref{sec:background}, we review the basics of cavity QED, introduce the formalism to treat light-matter interaction in atom arrays, and describe our setup. In \secref{sec:atomcav}, we analyze in detail the atom-array cavity and compute the cavity QED parameters $g$, $\kappa$, and $\gamma_\mathrm{3D}$. We address practical concerns, including transmission through the cavity and the effect of motion, in \secref{sec:experiment}. The section further includes a scheme to achieve a high cooperativity by applying a position-dependent Stark shift to flat mirrors instead of curving them. We conclude in \secref{sec:conclusion}, providing an outlook on the future potential of atom-array cavities.

\section{Background and setup\label{sec:background}}

\subsection{Cavity QED}\label{sec:cavityQED}

In this section, we review the aspects of cavity QED pertinent to our work. We refer the reader to review articles for a more comprehensive discussion~\cite{walther2006cavity,reiserer2015cavity}.

The interaction between a two-level atom and a single cavity mode within the rotating-wave approximation is described by the Jaynes-Cummings Hamiltonian
\begin{equation}\label{eq:cavity QED}
    \hat{H}_\mathrm{cQED} = \omega_\mathrm{a} \hat{\sigma}^+ \hat{\sigma}^- + \omega_\mathrm{c} \hat{a}^\dagger \hat{a} - g \left( \hat{\sigma}^+ \hat{a} +  \hat{\sigma}^- \hat{a}^\dagger \right),
\end{equation}
where we set $\hbar = 1$. Here, $\hat{\sigma}^\pm$ are the raising and lowering operators of the atom and $\hat{a}$ is the photon annihilation operator. The frequencies $\omega_\mathrm{a}$ and $\omega_\mathrm{c}$ refer to the resonant frequencies of the atom and the cavity, respectively. The coupling strength $g$ is given by $g = \vec{d} \cdot \vec{E}_0 / \hbar$, where $\vec{d}$ is the transition dipole moment of the atom and $\vec{E}_0$ is the electric field due to a single photon at the location of the atom. The magnitude of the single-photon field can be expressed as $|\vec{E}_0| = \sqrt{\hbar \omega_\mathrm{c} / 2 \epsilon_0 V} u(\vec{r}_\mathrm{a})$. Here, $u(\vec{r}_\mathrm{a})$ is the mode function of the electric field at the location of the atom, normalized to $1$ at the field maximum, and $V = \int \di^3 \vec{r} \, u^2(\vec{r})$ is the cavity mode volume.

In addition to the coherent Hamiltonian dynamics, cavity QED systems are also subject to incoherent processes such as spontaneous emission and decay of the cavity field. In free space, the spontaneous emission rate of a two-level system is given by $\gamma_\mathrm{a}=|\vec{d}|^2 \omega_\mathrm{a}^3/ 3\pi\epsilon_0\hbar c^3$. The presence of the cavity modifies this decay rate to $\gamma_\mathrm{3D}$, where the subscript highlights that the rate is associated with emission into unconfined, three-dimensional modes as opposed to cavity-mediated decay. If the cavity is composed of two mirrors with transmission coefficient $\mathcal{R}$ separated by a distance $L$, then the cavity field decays at a rate $\kappa = (1 - \mathcal{R}) c / L$.

The performance of a cavity QED setup is governed by the strength of the coherent interaction, $g$, compared to the incoherent rates, $\kappa$ and $\gamma_\mathrm{3D}$. The relative strength of the coherent to incoherent rates is captured by the cooperativity
\begin{equation}\label{eq:coop}
    C=\frac{4g^2}{\kappa\gamma_\mathrm{3D}} .
\end{equation}
We highlight that the cooperativity is independent of the properties of the atom, i.e.~the transition dipole moment and the transition frequency, when $\omega_\mathrm{a} / \omega_\mathrm{c} \approx 1$, which is the relevant regime for optical transitions.

\subsection{Collective light-matter interaction\label{sec:light-matter}}
Our setup consists of $N$ atoms located at positions $\vec{r}_i$, where $i \in \{1, 2, \ldots, N \}$ labels the atoms. For simplicity, we focus on two-level systems described by raising and lowering operators $\sigma_i^\pm$, although formalism can be readily generalized to more complex level schemes. We require that the transition frequencies $\omega_i$ are all comparable in the sense that $| \omega_i - \omega_j| \ll (\omega_i + \omega_j) / 2$. This allows us to replace $\omega_i$ by a typical transition frequency $\omega_0$ in several places, including the free space emission rates $\gamma_i = |\vec{d}_i|^2 \omega_0^3/ 3\pi\epsilon_0\hbar c^3$, where $\vec{d}_i$ is the transition dipole moment of atom $i$. For future reference, we further define $k_0 = \omega_0 / c$ and $\lambda_0 = 2 \pi c / \omega_0$.

We assume that the atoms are subject to an external driving field with optical frequency $\omega_\mathrm{L}$ and slowly varying, spatially dependent envelope $\vec{E}_0(\vec{r}, t)$, which results in the local Rabi frequencies $\Omega_i(t) = \vec{d}_i^* \cdot \vec{E}_0(\vec{r}_i, t)$. In a quantized treatment of the light, $\vec{E}_0(\vec{r}, t)$ should be identified with the expectation value of the positive frequency component $\hat{\vec{E}}_0^+(\vec{r}, t)$ of the electric field of a coherent state. For computational purposes, however, it is often most convenient to trace out the photons. Following standard formalism~\cite{asenjo2017exponential} within the assumptions described in Appendix~\ref{ap:effHam}, we arrive at the effective non-Hermitian Hamiltonian
\begin{equation}\label{eq:drivenHam}
\begin{split}
    \hat{H} = &\sum_{i, j} \left[ (\omega_i - \omega_\mathrm{L}) \delta_{ij} + \Delta_{ij} - \frac{i}{2} \Gamma_{ij} \right] \hat{\sigma}_i^+ \hat{\sigma}_j^- \\
    & - \sum_i \left( {\Omega}_i\hat{\sigma}_i^+ +\mathrm{h.c.} \right) ,
\end{split}
\end{equation}
where we work in the frame that rotates with frequency $\omega_\mathrm{L}$. The coefficients $\Delta_{ij}$ and $\Gamma_{ij}$ are Hermitian matrices that capture the coherent and dissipative interaction between atoms mediated by the electromagnetic field. An explicit expression in terms of the electromagnetic Green's function is provided in Appendix~\ref{ap:effHam}.

To fully describe the dynamics of the atoms, the non-Hermitian Hamiltonian in \eqref{eq:drivenHam} must be supplemented by quantum jump terms~\cite{gardiner2004quantum}. In the limit of weak driving, however, the effect of the quantum jumps is subleading and can be neglected. Moreover, in this regime, the number of excited atoms remains small, which allows us to treat the atomic spin operators, $\hat{\sigma}^\pm_i$, as bosonic operators. We work within these approximations throughout this paper unless stated otherwise.

For many applications, we are not only interested in the internal dynamics of the atoms but also in the scattered field. It is possible to reconstruct the electric field using the general input-output relation~\cite{gardiner1985input,asenjo2017exponential}
\begin{equation}\label{eq:IO}
    \hat{\vec{E}}^+(\vec{r}, t) = \hat{\vec{E}}_0^+(\vec{r}, t) + \frac{1}{\epsilon_0} \frac{\omega_0^2}{c^2} \sum_i \vec{G}(\vec{r}, \vec{r}_i; \omega_0) \cdot \vec{d}_i \, \hat{\sigma}_i^-(t) ,
\end{equation}
where the operator $\hat{\vec{E}}^+(\vec{r}, t)$ describes the slowly varying envelope of the positive frequency component of the electric field in the Heisenberg-Langevin picture.

\subsection{Atom arrays as resonant mirrors}\label{sec:atommirror}
To illustrate the above formalism, we review the reflection from a two-dimensional array of atoms. We first consider an infinite, regular lattice of identical atoms with transition frequency $\omega_0$ and linearly polarized transition dipole moment $\vec{d}_0$. The Hermitian terms $\Delta_{ij}$ describe long-ranged hopping of excitations, which leads to a nontrivial dispersion relation. Similarly, the anti-Hermitian terms $i \Gamma_{ij}$ correspond to a dissipative coupling that enhances or suppresses the spontaneous emission of delocalized excitations, which results in bright and dark states. Owing to the translational symmetry of the infinite lattice, both of these matrices are diagonal in momentum space. By computing the discrete Fourier transform of $\Delta_{ij}$ and $\Gamma_{ij}$, we find that the Bloch mode with crystal momentum $\vec{k}$, created by $\hat{\sigma}^+_{\vec{k}} = \sum_i \hat{\sigma}_i^+ e^{- i \vec{k} \cdot\vec{r}_i} / \sqrt{N}$, is detuned from the laser by $\Delta(\vec{k})$ and decays at a rate $\Gamma(\vec{k})$~\cite{shahmoon2017cooperative}.

To study the scattering properties of the array, we solve the dynamics under \eqref{eq:drivenHam} with a weak, monochromatic drive to obtain the steady state $\langle \sigma_\vec{k}^- \rangle = \Omega(\vec{k}) / [ \Delta(\vec{k}) - i \Gamma(\vec{k}) / 2 ]$, where $\Omega(\vec{k})$ is the discrete Fourier transform of $\Omega_i$. Using \eqref{eq:IO}, we find that the amplitude reflection coefficient for an incident plane wave with in-plane momentum $\vec{k}$ is given by
\begin{equation}\label{eq:r1mirror}
    r(\vec{k}, \omega_\mathrm{L}) = \frac{i \Gamma({\vec{k}}) / 2}{\omega_0 - \omega_\mathrm{L} + \Delta(\vec{k}) - i \Gamma({\vec{k}}) / 2} .
\end{equation}
This expression holds if the lattice spacing is sufficiently small such that higher-order Bragg scattering is suppressed. We further assumed that the projection of the electric field onto the plane of the array is parallel to the transition dipole moment. The transmission coefficient is given by $t(\vec{k}, \omega_\mathrm{L}) = 1 + r(\vec{k}, \omega_\mathrm{L})$. We will frequently work with the intensity reflection and transmission coefficients $\mathcal{R} = |r|^2$ and $\mathcal{T} = |t|^2$, which by energy conservation satisfy $\mathcal{R} + \mathcal{T} = 1$ for an infinite array. It is always possible to tune the laser on resonance with a particular Bloch mode, $\omega_\mathrm{L} = \omega_0 + \Delta(\vec{k})$, such that $r(\vec{k}, \omega_\mathrm{L}) = -1$ and $t(\vec{k}, \omega_\mathrm{L}) = 0$. Atom arrays with subwavelength lattice spacings are thus excellent mirrors, albeit only over a narrow frequency range of width $\Gamma(\vec{k})$ centered around the collective resonance. At normal incidence ($\vec{k} = 0$) and in the absence of Bragg scattering, the linewidth is given by $\Gamma(0) = 3\pi \gamma_0 / k_0^2 A$, where $A$ is the area of the unit cell of the array~\cite{shahmoon2017cooperative}.

\subsection{Array cavities}
We now introduce the specifics of the setup considered in this work. The cavity is formed by two atom arrays separated by a distance $L$, where each array consists of a square lattice of $N\times N$ atoms with lattice spacing $a$ as depicted in Fig.~\ref{figIntro}(a). One or more additional atoms are placed inside the cavity. We refer to the atoms forming the arrays as ``array atoms'' and to the atoms inside the cavity as ``target atoms''.

We model all atoms as two-level systems with a transition dipole moment that is linearly polarized along the direction of one of the lattice vectors. We assume that the array atoms are identical with resonance frequency $\omega_0$ and free-space decay rate $\gamma_0$. The corresponding quantities of the target atoms, denoted by $\omega_\mathrm{a}$ and $\gamma_\mathrm{a}$, can in general be different. In the applications considered below, the resonance frequencies $\omega_0$ and $\omega_\mathrm{a}$ differ by an amount proportional to $\gamma_0$ and we require $\gamma_\mathrm{a} \ll \gamma_0$. The target and array atoms may nevertheless belong to the same species as the transition frequency and decay rate can be tuned by local dressing fields as described in \secref{sec:protocols}.

The wavelength $\lambda_0 = 2 \pi c / \omega_0$ is an important length scale of the system. We will neglect the difference between $\lambda_0$ and $\lambda_\mathrm{a} = 2 \pi c / \omega_\mathrm{a}$ throughout, which is valid because $|\omega_0-\omega_\mathrm{a}|\ll\omega_0$. For the arrays to act as resonant mirrors without Bragg scattering at normal incidence, we require subwavelength lattice spacing, $a < \lambda_0$. We treat $a$ as a free parameter and discuss how the cavity QED parameters depend on it. We showcase many of our results at a fixed lattice spacing of $a=0.47\lambda_0$, which corresponds to a predicted magic wavelength for the $D_2$ manifold of $^{87}\mathrm{Rb}$~\cite{arora2007magic}. To achieve high quality factors, it is necessary to curve the arrays such that they match the optical wavefront of a Gaussian beam with waist radius $w_0$. To this end, we shift the lattice atoms in the direction of the cavity axis, as detailed in Appendix~\ref{ap:phasematching}.

An important quantity in our analysis will be the ratio
\begin{equation}
    \zeta = \frac{\Gamma_0 L}{2 c},
\end{equation}
where $\Gamma_0$ is the linewidth of a single array. As discussed in the introduction, $\zeta$ is the ratio of the propagation time $\tau_\mathrm{prop} = L / c$ to the Wigner time delay $\tau_\mathrm{delay} = 2 / \Gamma_0$. Within the paraxial limit, we expect $\Gamma_0$ to be close to the decay rate of the zero momentum of an infinite array. From \secref{sec:atommirror}, we find $\Gamma_0 \approx 3 \pi \gamma_0 / (k_0 a)^2$ and
\begin{equation}
    \label{eq:zeta_approx}
    \zeta \approx \frac{3 \pi}{2} \frac{1}{(k_0 a)^2} \frac{\gamma_0 L}{c} .
\end{equation}
We will assume throughout that $\gamma_0 L / c \ll 1$ and $\zeta \ll 1$, which follows provided $a$ is not deeply subwavelength. These assumptions are valid for optical dipole transitions and array cavities with lengths on the order of a few wavelengths. The condition $\gamma_0 L / c \ll 1$ indicates that retardation is negligible, which is separately required to apply the formalism of \secref{sec:light-matter}.

\section{Atom-cavity coupling}\label{sec:atomcav}

\begin{figure}[tb]
	\includegraphics[width=0.99\linewidth]{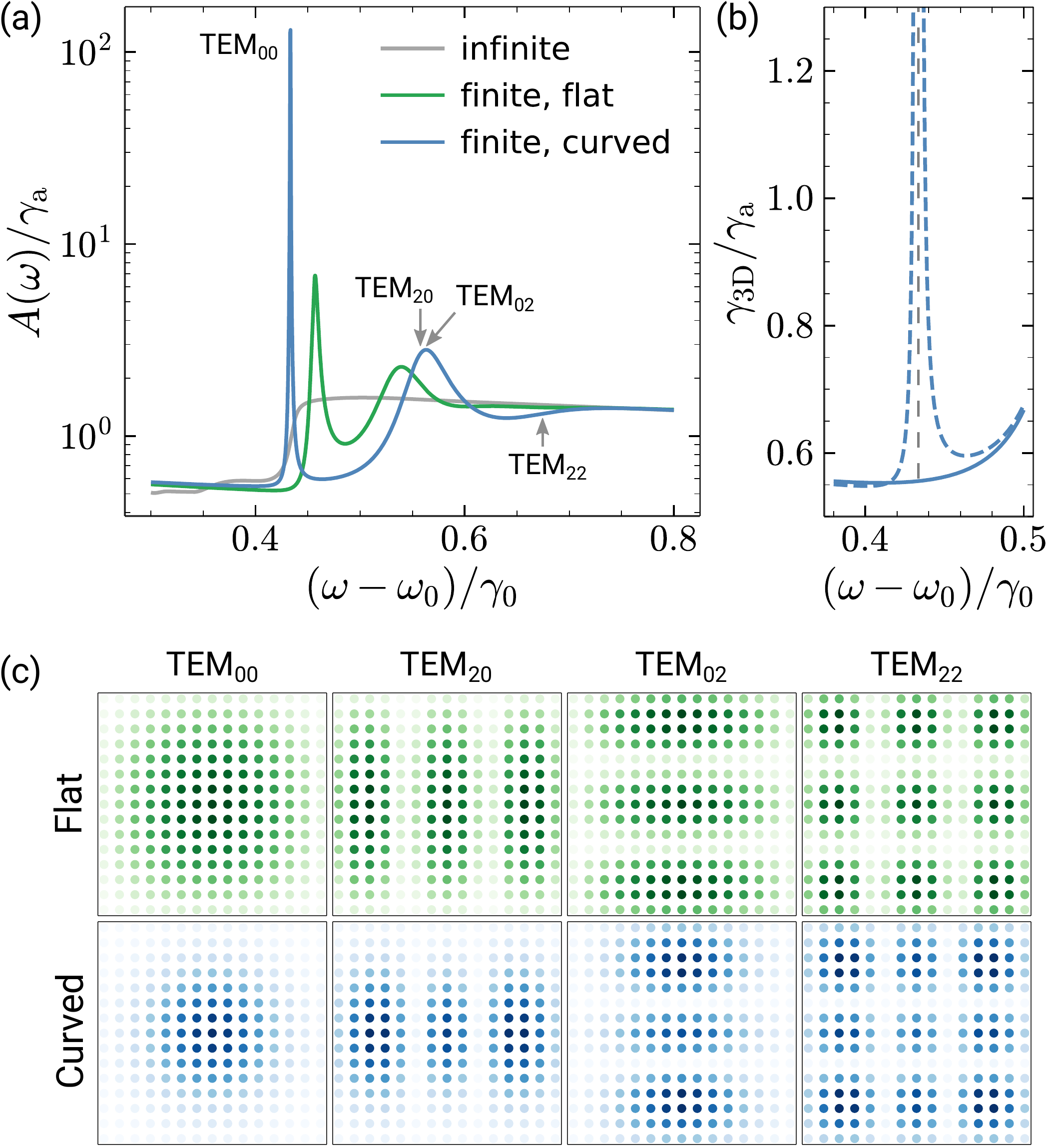}
    \caption{\label{figAtomCav} Spectral response of a target atom placed at the center of an array cavity. (a)~Spectral function, \eqref{eq:response}, for a square lattice arrays with lattice spacing $a = 0.47 \lambda_0$ and distance $L = 1.5 \lambda_0$ between the arrays. We consider infinite arrays and finite arrays composed of $N \times N = 15 \times 15$ atoms. For the blue line, the arrays are curved to match a Gaussian beam with waist $w_0 = 2 \lambda_0$ and the labels indicate the energy of the eigenmodes illustrated in (c).
    (b)~By subtracting the contribution of the TEM$_{00}$ mode from the spectral function close to the cavity resonance (dashed curve), we obtain a smooth background (solid curve). The value of the background at the cavity resonance can be interpreted as $\gamma_\mathrm{3D}$, i.e., the modified decay rate of the target atom into free space.
    (c)~Illustration of the four eigenmodes of $\mathbf{H}^\mathrm{AA}$ with the largest contribution to the spectral function for finite arrays with the same parameters as in (a). For each atom in one of the arrays, we draw a circle whose color is determined by the absolute value of the amplitude that the atom is excited. Darker colors correspond to larger amplitudes.
	}
\end{figure}

\subsection{Cavity modes}\label{sec:cavitymodes}
To formalize the connection between cavity QED and the above setup, we consider the dynamics of a single target atom initialized in the excited state at time $t = 0$. The array atoms all start in the ground state. As shown in Appendix~\ref{ap:selfenergy}, the amplitude that the excitation is on the target atom after time $t$ is given by
\begin{equation}
    \label{eq:ca}
    c_\mathrm{a}(t) = \frac{i}{2 \pi} \int_{-\infty}^\infty \di \omega \, e^{-i \omega t} \frac{1}{\omega-\omega_\mathrm{a}+ i \gamma_\mathrm{a} / 2 - \Sigma_\mathrm{a}(\omega)}\, .
\end{equation}
The function $\Sigma_\mathrm{a}(\omega)$, known as the self-energy, describes a frequency-dependent modification of the resonance frequency and decay rate of the target atom due to the presence of the arrays. We denote by $\ket{i}$ the state in which the atom with index $i$ is excited and all other atoms are in the ground state. We choose $i = 0$ for the target atom and $i \in \{1, 2, \ldots, 2 N^2 \}$ for the array atoms. We define the $2 N^2$ vector components $H^{\mathrm{TA}}_{i} = \braket{0 | \hat{H} | i}$ and $H^{\mathrm{AT}}_{i} = \braket{i | \hat{H} | 0}$ and the $2 N^2 \times 2 N^2$ matrix $H^{\mathrm{AA}}_{ij} = \braket{i | \hat{H} | j}$, with $i$ and $j$ restricted to array atoms. The self-energy may then be written as 
\begin{equation}\label{eq:selfE}
    \Sigma_\mathrm{a}(\omega) = \vec{H}^{\mathrm{TA}} \cdot \left[\omega \vec{I} - \vec{H}^\mathrm{AA} \right]^{-1} \cdot \vec{H}^{\mathrm{AT}},
\end{equation}
where $\vec{I}$ is the $2 N^2 \times 2 N^2$ identity matrix.

Since the self-energy is complex, it is convenient to consider its real and imaginary parts separately. We focus in particular on the spectral function
\begin{equation}\label{eq:response}
    A(\omega) = \gamma_\mathrm{a} - 2\, \mathrm{Im} [ \Sigma_\mathrm{a}(\omega) ] \, ,
\end{equation}
which can in certain regimes be interpreted as the modified decay rate of the target atom due to the presence of the arrays. To see this, we observe that if $\Sigma(\omega)$ varies slowly, the target atom experiences the array atoms as a bath that can be treated within the Markov approximation. Formally, this corresponds to replacing the self-energy by a constant in \eqref{eq:ca}, resulting in the exponential dependence $c_a(t) \approx \exp \{ -i [ \omega_a - i \gamma_a / 2 + \Sigma(\omega_a) ] t \}$. This interpretation is valid provided that the self-energy is approximately constant over the frequency interval $\omega_a \pm |\Sigma(\omega_a)|$. Because $\Sigma(\omega)$ is proportional to $\gamma_a$, this condition can always be satisfied for smooth $\Sigma(\omega)$ and a sufficiently small $\gamma_a$.

In Fig.~\ref{figAtomCav}(a), we plot $A(\omega)$ for a target atom placed at the center of two arrays separated by $L = 1.5 \lambda_0$. We first consider a pair of infinite square lattices, for which we compute the self-energy numerically by converting the sum over lattice sites into an integral over momenta~\cite{shahmoon2017cooperative}. The spectral function displays plateaus, where the emission rate of the target atom is either suppressed or enhanced compared to free space. The plateaus are the result of coupling to a continuum of guided modes supported by the arrays. The sharp transitions between the plateaus arise at the band edges of the dispersion relation of these modes. This behavior is in stark contrast to finite-sized arrays, for which $A(\omega)$ displays sharp resonances on top of a smooth background. Due to the nearly singular behavior of the spectral energy at these resonances, they cannot be interpreted as a simple modification of the decay rate.

The form of \eqref{eq:selfE} indicates that the resonances arise from coupling of the target atom to individual eigenmodes of the array. Concretely, an eigenmode of $\vec{H}^\mathrm{AA}$ with complex eigenvalue $\omega_c - i \kappa/2$ and associated left and right eigenvectors $\vec{v}^\mathrm{L}$ and $\vec{v}^\mathrm{R}$ contributes $g^2 / (\omega - \omega_c + i \kappa / 2)$ to the self-energy, where $g^2 = (\vec{H}^\mathrm{TA} \cdot \vec{v}^\mathrm{R}) (\vec{v}^\mathrm{L} \cdot \vec{H}^\mathrm{AT})$ quantifies the coupling strength between the eigenmode and the target atom. Hence, an eigenmode with small decay rate $\kappa$ and large coupling strength $g$ gives rise to a sharp resonance of height $g^2 / \kappa$. The resonance takes the same shape as the resonance found in the atom self-energy of the Jaynes-Cummings model with corresponding parameters (see Appendix~\ref{ap:selfenergy} for details). We can thus view the eigenmodes that lead to resonances in the self-energy as cavity modes.

To characterize the cavity modes, we numerically diagonalize the matrix $\vec{H}^\mathrm{AA}$ and compute the ratio $g^2 / \kappa$ for each eigenmode. The distribution of the excitation across the array according to the right eigenvector is illustrated in Fig.~\ref{figAtomCav}(c) for the eigenmodes with the four largest magnitudes of $g^2 / \kappa$. The distributions closely resemble the profile of Hermite-Gaussian modes, $\mathrm{TEM}_{mn}$. We identify the sharpest resonance in the spectral function with the fundamental $\mathrm{TEM}_{00}$ mode and the second largest peak with the degenerate $\mathrm{TEM}_{20}$ and $\mathrm{TEM}_{02}$ modes. The next smaller value of $g^2/\kappa$, corresponding to the $\mathrm{TEM}_{22}$ mode, is insufficient to give rise to a discernible resonance. We note, however, that higher-order resonances can be observed in larger arrays. Odd modes, such as $\mathrm{TEM}_{10}$, do not appear in Fig.~\ref{figAtomCav}(a) because they do not couple to the target atom.

We observe that the frequency splitting between the $\mathrm{TEM}_{00}$ and $\mathrm{TEM}_{mn}$ modes, denoted by $\delta_{mn}$, is on the order of $\gamma_0$. This frequency scale, set by the array atoms, is not present in a conventional cavity, where the mode splitting is instead given by $\delta_{mn}^\mathrm{conv} = (c / L)(m + n) \arccos (1 - L/R)$. Here, $R = (L/2) + (k_0^2 w_0^4 / 2 L)$ is the radius of curvature of the mirrors. The expression for $\delta_{mn}^\mathrm{conv}$ can be obtained by matching the propagation phase to the Gouy phase~\cite{siegman1986lasers}. The transverse mode splitting in a conventional cavity is therefore proportional to the free spectral range, which is much greater than $\gamma_0$ for the range of parameters considered here. The mode splitting in the array cavity differs due to the dispersive response of the arrays. According to \eqref{eq:r1mirror}, reflection by one of the mirrors results in a phase shift $\phi = \arctan(2 \delta_{mn} / \Gamma_0)$, where we assumed that the fundamental mode is resonant with the mirror. In the regime $\zeta \ll 1$, this phase shift is much greater than the propagation phase. We therefore replace the propagation phase in the resonance condition by $\phi$ to obtain
\begin{equation}
    \label{eq:mode_splitting}
    \delta_{mn} \approx \frac{\Gamma_0}{2} \tan \left[ (m+n) \arccos \left( 1 - \frac{L}{R} \right) \right] .
\end{equation}
For the finite, curved arrays in \figref{figAtomCav}(a), this expression yields $\delta_{20} \approx 0.13 \, \gamma_0$, in perfect agreement with the numerically computed splitting of the resonances in the spectral function.

We note that the above expression only applies if $\delta_{mn} \ll \Gamma_0$ since the arrays otherwise reflect only weakly and do not give rise to a cavity resonance. Moreover, there are significant deviations from \eqref{eq:mode_splitting} in small arrays with little or no curvature. This is evident for the finite, flat arrays in \figref{figAtomCav}(a), where the TEM$_{00}$ and TEM$_{20}$ modes are clearly split, although they should be degenerate according to \eqref{eq:mode_splitting}. The observed mode splitting is due to boundary effects, which effectively localize the beam to a smaller waist. The boundaries also induce diffraction losses and thereby broaden the resonances~\cite{manzoni2018optimization}.

Below, we focus on the fundamental mode, $\mathrm{TEM}_{00}$, which couples most strongly to the target atom.  This is justified if the target atom is tuned close to the resonance of this mode and $\gamma_\mathrm{a}$ is small enough such that off-resonant coupling to other modes is negligible. We will treat the cavity mode separately from all other eigenmodes and may therefore subtract its contribution from the self-energy to recover a smooth frequency dependence as shown in Fig.~\ref{figAtomCav}(b). This background value can be interpreted as the modification of the properties of the target atom by the weakly coupled eigenmodes and allows us to compute the modified decay rate $\gamma_\mathrm{3D}$ into free space. Before proceeding to applications, we quantitatively analyze the dependence of the cavity QED parameters $g$, $\kappa$, $\gamma_\mathrm{3D}$, and the cooperativity $C$ on the design parameters $w_0$, $L$, and $a$.

\subsection{Cavity parameters}\label{sec:parameters}
\subsubsection{Coupling strength, $g$}\label{sec:g}

As noted above, a particular eigenmode of the array with left and right eigenvector $\vec{v}^\mathrm{L}$ and $\vec{v}^\mathrm{R}$ couples to the target atom with strength $g = \sqrt{ (\vec{H}^\mathrm{TA} \cdot \vec{v}^\mathrm{R}) (\vec{v}^\mathrm{L} \cdot \vec{H}^\mathrm{AT}) }$. Since the Hamiltonian is non-Hermitian, $g$ will in general be complex. Its imaginary part captures the fact that the field emitted by the target atom and the eigenmode of the array interfere, which leads to enhanced or suppressed collective emission. However, in the regimes discussed in this work, the imaginary part of $g$ is much smaller than its real part. We will therefore neglect the imaginary part throughout.

\begin{figure}[t]
	\includegraphics[width=\linewidth]{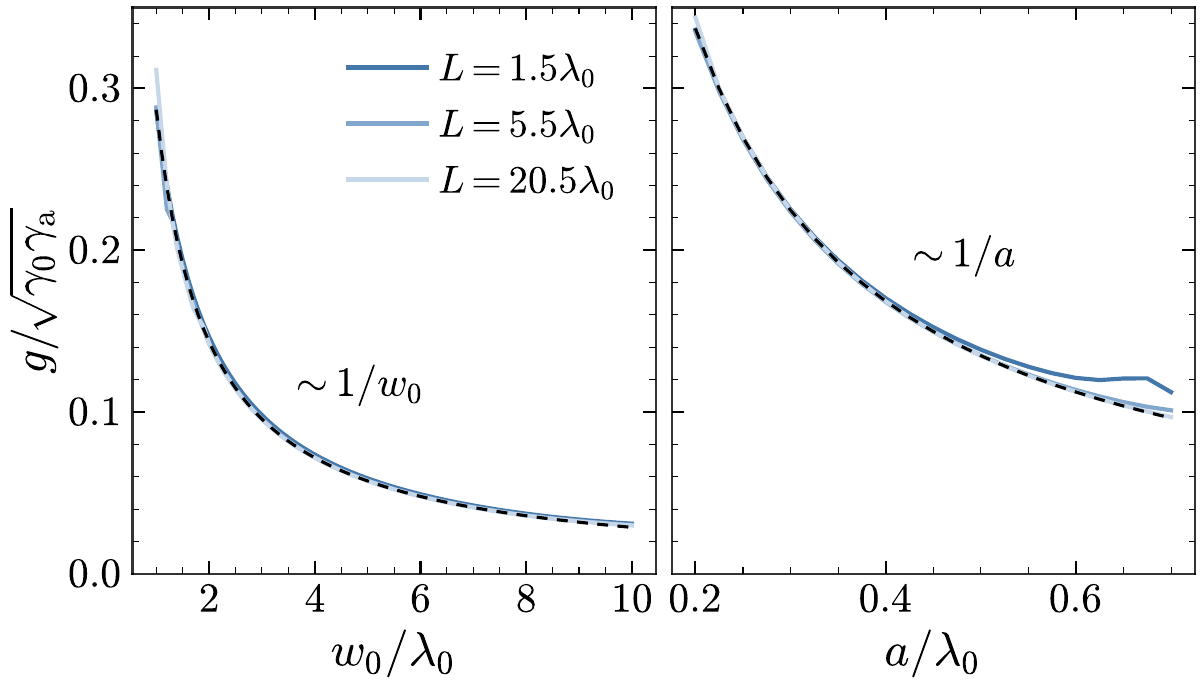}
    \vspace{-10pt}
	\caption{\label{figCoupStrength} Coupling strength, $g$, between the fundamental mode of an array cavity and a target atom placed at the field maximum. The left panel shows the dependence on the beam waist $w_0$, which is determined by the curvature of the mirrors, for a fixed lattice spacing $a = 0.47 \lambda_0$ and array size $N\times N=60\times 60$. In the right panel, we vary $a$ with $w_0 = 2 \lambda_0$ fixed and $N\times N=45\times 45$. The three curves for different separations of the array mirrors (legend in the left panel) lie on top of each other except for the largest values of $a$. The dashed lines represent the analytical prediction, \eqref{eq:g}.	
 }
\end{figure}

In Fig.~\ref{figCoupStrength}, we show the coupling strength $g$ computed in this way as a function of the lattice spacing $a$, the cavity length $L$, and the beam waist $w_0$. These parameters determine the radius of curvature of the array mirrors via the relation $R = (L/2) + (k_0^2w_0^4/ 2 L)$. We observe that $g$ is inversely proportional to $w_0$ and $a$ but largely independent of $L$. To interpret this result, we compare it to the case of a conventional cavity. According to \secref{sec:cavityQED}, the maximum coupling strength in a conventional cavity can be written as
\begin{equation}\label{eq:gconv}
    g^\mathrm{conv} = \sqrt{\frac{3\pi}{2}\,\frac{\gamma_\mathrm{a}\omega_0}{k_0^3 V}}\,,
\end{equation}
where the mode volume for a Gaussian beam is given by $V= \pi w_0^2 L / 4$~\cite{tanji-suzuki2011interaction}. Although this expression correctly captures the dependence of $g$ on $w_0$, it is clearly inconsistent with the observed dependence on $L$ and $a$.

Even though we computed the value of $g$ from the spin model in \eqref{eq:drivenHam}, its physical origin lies in the coupling of the target atom to the electric field of a single excitation in the cavity mode. The reason for the discrepancy can then be understood from the fact that \eqref{eq:gconv} does not include the time delay $\tau_\mathrm{delay}$ in the reflection caused by the finite bandwidth of the array mirrors. Due to the delay, the field tends to localize close to the arrays and we expect that the energy density at the center of the cavity will be suppressed by the factor $\zeta$. Since the target atom couples to the electric field, which is proportional to the square root of the energy density, the argument suggests that the coupling strength in an array cavity is given by
\begin{equation}\label{eq:g}
    g \approx \sqrt{\zeta}\cdot g^\mathrm{conv} \approx \frac{\sqrt{9 \pi \gamma_\mathrm{a} \gamma_0}}{k_0^2w_0a} \, ,
\end{equation}
where we used \eqref{eq:zeta_approx}. This expression correctly captures the dependence on $w_0$, $L$, and $a$ observed in \figref{figCoupStrength}. It is indeed in excellent quantitative agreement over a wide range of parameters as shown by the black dashed curve. The only significant deviations occur for large values of $a / \lambda_0$ and small $L$, predominantly $L=1.5\lambda_0$, which we attribute to near-field coupling between the target atom and the arrays.

\begin{figure*}[t]
	\includegraphics[width=\linewidth]{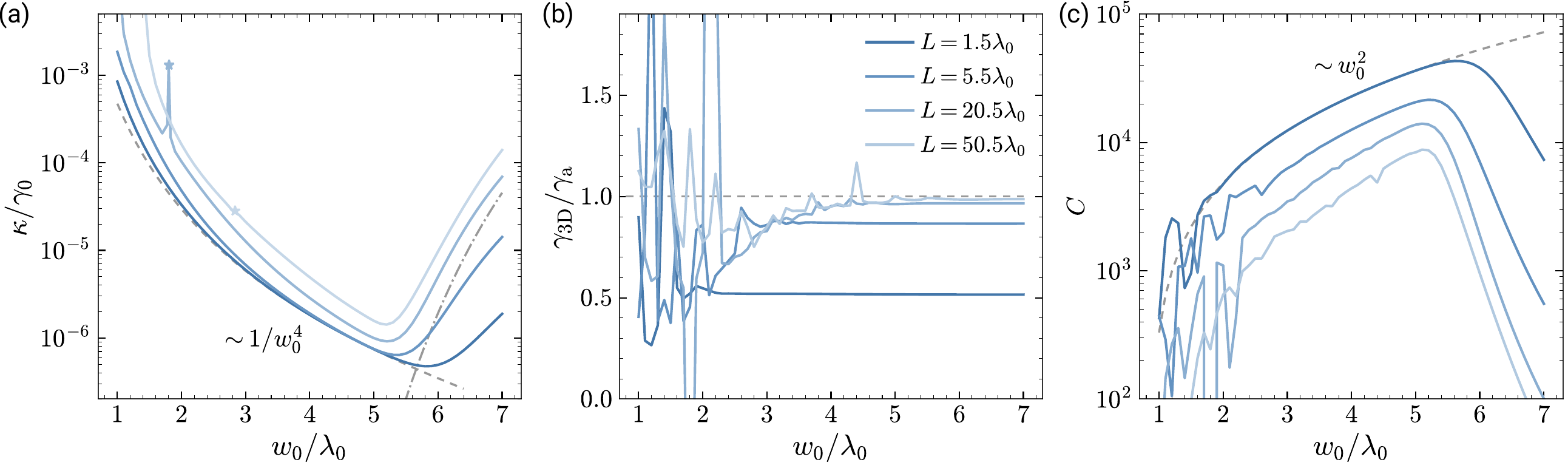}
	\caption{\label{fig_gkaC} Cavity decay rate, $\kappa$, decay rate of the target atom, $\gamma_\mathrm{3D}$, and cooperativity, $C$, for a target atom placed at the maximum of intensity. All panels show the dependence on the beam waist $w_0$, determined by the curvature of the mirrors, for a fixed lattice spacing $a=0.47\lambda_0$ and array size $N \times N = 60 \times 60$. We include four different values of $L$ (legend in the central panel). (a)~The dashed-dotted line represents the prediction of \eqref{eq:kappa} using an estimate for the portion of the beam that extends beyond the mirrors, $1 - \mathcal{R} \approx \mathrm{Erf}^2 (N a / \sqrt{2} w_0)$. The dashed line shows the scaling $\kappa \sim 1/w_0^4$ arising from nonzero in-plane momentum components. The stars indicate the values of $w_0$ for which the cavity is in the confocal configuration, $R=L$. At this point, the fundamental mode becomes degenerate with higher-order $\mathrm{TEM}_{mn}$ modes, which may lead to increased loss.
    (b)~The values of $\gamma_\mathrm{3D}$ do not follow a smooth curve at small beam waists due to the presence of nearby cavity modes [see \eqref{eq:mode_splitting}]. In this regime, $\gamma_\mathrm{3D}$ cannot be simply interpreted as a decay rate.
    (c)~The cooperativity follows the relation $C \sim w_0^2$ (dashed curve) up until the reflection coefficient is significantly affected by the finite size of the mirrors.
 }
\end{figure*}

\subsubsection{Cavity decay rate, $\kappa$}\label{sec:kappa}
Following our discussion of the self-energy, the cavity decay rate $\kappa$ is determined by the imaginary part of the eigenvalue of the non-Hermitian Hamiltonian $\vec{H}^\mathrm{AA}$ belonging to the array eigenmode that is responsible for the cavity resonance. In Fig.~\ref{fig_gkaC}(a) we show the cavity linewidth as a function of $w_0$ and $L$ obtained by numerically diagonalizing $\vec{H}^\mathrm{AA}$ for an array cavity of size $N \times N = 60 \times 60$. We observe the smallest decay rate when the beam waist is a few wavelengths, at which point $\kappa < 10^{-6} \gamma_0$ can be achieved.

To explain the qualitative features in \figref{fig_gkaC}(a), we again appeal to a physical picture involving the electric field, despite it having been traced out in our calculation. In a conventional cavity comprising identical mirrors with reflection coefficient $\mathcal{R}$, a fraction $1 - \mathcal{R}$ of the energy is lost at each reflection, leading to the cavity decay rate $\kappa^\mathrm{conv} = c (1 - \mathcal{R}) / L$. For the array mirrors, we have to take into account that the reflection time delay is much greater than the round-trip time. Therefore, we expect
\begin{equation}
    \label{eq:kappa}
    \kappa \approx \zeta \kappa^\mathrm{conv} = (1 - \mathcal{R}) \frac{\Gamma_0}{2} \, .
\end{equation}
Since $\Gamma_0$ is approximately independent of $w_0$ and $L$, the dependence of $\kappa$ on these parameters is governed by the reflection coefficient of a single mirror. We outline the pertinent features of the reflection coefficient below, while postponing a more detailed analysis to the discussion of the cavity transmission in \secref{sec:transmission}.

If the beam waist is large, reflection is imperfect because the arrays do not capture the entire beam. This effect was analyzed in detail in Ref.~\cite{manzoni2018optimization} for a Gaussian beam focused onto a flat array, where it reduces the reflection coefficient to $1 - \mathcal{R} \approx \mathrm{Erf}^2 (N a / \sqrt{2} w_0)$. The expression qualitatively captures the increase in $\kappa$ at large values of $w_0$ [see dashed-dotted line \figref{fig_gkaC}(a)]. There are however significant quantitative deviations, that can be attributed to the dependence of diffraction losses on the separation between the mirrors, as diffracted light spreads out over longer distances between reflections~\cite{li1965diffraction}, and to the modification of the cavity mode due to boundary effects, as discussed in relation to \figref{figAtomCav}(c). We note that the finite size of the arrays also leads to losses at small values of $w_0$, especially when $L$ is large, owing to the strong divergence of the cavity mode.

In the regime where the arrays are sufficently large to contain the cavity mode, we observe that $\kappa$ is roughly proportional to $1/w_0^4$ [dashed line in \figref{fig_gkaC}(a)]. This scaling was shown in Ref.~\cite{manzoni2018optimization} to be the result of higher in-plane momentum components of the Gaussian beam, which are imperfectly reflected due to the quadratic dispersion relation of the eigenmodes of the array. The value of $\kappa$ is independent of the size of the array in this regime. We conclude that to minimize $\kappa$, one should choose the largest possible value of $w_0$ for which losses due to the finite size of the array are negligible. We note that other decay channels may significantly modify the dependence of $\kappa$ on $w_0$. In \secref{sec:transmission}, we consider enhanced cavity decay by detuning the cavity resonance from the array mirrors, while in \secref{sec:motion} we explore the detrimental contributions due to motion and disorder.

\subsubsection{Decay rate of the target atom, $\gamma_\mathrm{3D}$}\label{sec:gamma3D}

The presence of the array atoms modify the decay rate of the target atom. As discussed in the context of \figref{figAtomCav}(b), this decay rate, denoted by $\gamma_\mathrm{3D}$, can be computed by evaluating the spectral function after subtracting the contribution from the cavity mode. The resulting values are shown in Fig.~\ref{fig_gkaC}(b) as a function of $w_0$ and $L$.

For weakly curved mirrors (large $w_0$), $\gamma_\mathrm{3D}$ is close to the decay rate in free space, $\gamma_\mathrm{a}$, when the separation between the mirrors is much greater than the wavelength. The decay rate is reduced by about $50 \%$ for a short cavity with $L = 1.5 \lambda_0$, which can be interpreted as the suppression of free-space decay channels due to the modification of the electromagnetic environment caused by the mirror atoms. This effect tends to be enhanced as the cavity mode becomes more tightly focused. However, $\gamma_\mathrm{3D}$ stops being a smooth function for smaller $w_0$ and may even take unphysical negative values. We attribute this behavior to weak resonances with higher-order cavity modes that, according to \eqref{eq:mode_splitting}, can become degenerate with the fundamental mode. In such cases, the single-mode cavity assumption is not justified, and our method to compute $\gamma_\mathrm{3D}$ is not valid. These resonances also contribute to increased losses in \figref{fig_gkaC}(a).

\subsubsection{Cooperativity}
We now combine the results of the preceding sections to compute the cooperativity $C = 4 g^2 / \kappa \gamma_\mathrm{3D}$. From Eqs.~(\ref{eq:g}) and (\ref{eq:kappa}), we obtain
\begin{equation}
    \label{eq:c}
    C \approx \frac{6}{\pi^2} \frac{\gamma_\mathrm{a}}{\gamma_\mathrm{3D}} \frac{1}{1 - \mathcal{R}} \left( \frac{\lambda_0}{w_0} \right)^2 \, .
\end{equation}
We highlight that this expression also holds for conventional cavities because the factors of $\zeta$ in the expression for $g$ and $\kappa$ cancel. This reflects the fact that the cooperativity is independent of the round-trip time or, in the case of the array cavity, the lifetime of the mirror eigenstate associated with the cavity mode. Since the resonant cross section of the target atom is proportional to $\lambda_0^2$, the cooperativity has the appealing physical interpretation as the number of times that a cavity photon interacts with the atom during its lifetime~\cite{tanji-suzuki2011interaction}.  For a given reflection coefficient and curvature of the mirrors, we thus expect an array cavity and a conventional cavity to have approximately the same cooperativity. In practice, deviations may arise because the different setups lead to different values of $\gamma_\mathrm{3D}$.

As shown in Fig.~\ref{fig_gkaC}(c), the cooperativity reaches a maximum at the largest beam waist that is supported with low loss by the finite-sized arrays. For the parameters considered here, values exceeding $C = 10^4$ can be reached. Leading up to the maximum, the cooperativity is approximately proportional to $w_0^2$. This follows from the dependence $1 - \mathcal{R} \propto 1/w_0^4$, observed in \figref{fig_gkaC}(a), combined with the $1/w_0^2$ factor in \eqref{eq:c}. This result could be extrapolated to larger $w_0$ by increasing the size of the arrays to prevent the detrimental increase in $\kappa$. However, the splitting between the fundamental and higher $\mathrm{TEM}_{mn}$ modes will decrease for larger $w_0$ [see \eqref{eq:mode_splitting}], which may render the single-mode approximation invalid. Moreover, if other decay channels are present, $\mathcal{R}$ may become independent of $w_0$. In this case, we recover the more conventional scaling $C \propto 1/w_0^2$ and the cooperativity is maximized at the smallest achievable beam waist.

\begin{figure*}[t]
	\includegraphics[width=0.99\linewidth]{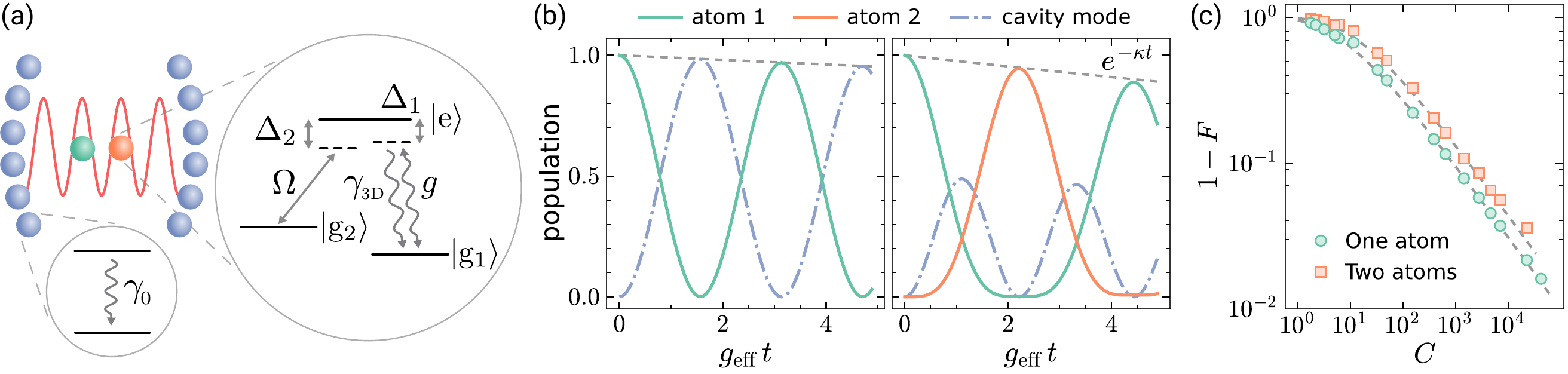}
	\caption{\label{figProt} Cavity QED dynamics in the strong coupling regime. (a)~To achieve strong coupling, a three-level $\Lambda$-system is introduced for the target atoms. Under a classical drive with large detuning $\Delta_2$, the excited state $\ket{\mathrm{e}}$ can be eliminated. The resulting effective coupling of the two-level system $\{ \ket{\mathrm{g}_1} , \ket{\mathrm{g}_2} \}$ to the cavity is controlled by the strength $\Omega$ of the drive. The array atoms remain unchanged.
    (b)~The left panel shows vacuum Rabi oscillations of a target atom placed at the maximum of field intensity of an array cavity with $a=0.47\lambda_0$, $L=1.5\lambda_0$, $w_0=5.5\lambda_0$, and $N\times N=60\times 60$. The right panel shows population exchange between two target atoms placed at the two adjacent field maxima near the center of a cavity with $a=0.47\lambda_0$, $L=4\lambda_0$, $w_0=3\lambda_0$ and $N\times N=45\times 45$. The setups correspond to cooperativities $C= 4.3\times 10^4$ and $C= 7.1\times 10^3$, respectively. We choose a large detuning $\Delta_2 = 500 \gamma_\mathrm{a}$ and set $\Delta_1 \approx \Delta_2$, where a small difference between the detunings corrects for Stark shifts in order to satisfy the two-photon resonance condition. In both panels, one of the atoms is initialized in the state $\ket{\mathrm{g}_2}$ before the system is evolved without approximations under the non-Hermitian Hamiltonian including the $\Lambda$-systems and all array atoms. We obtain the displayed cavity population by projecting the array atoms on the corresponding eigenstate of the $\vec{H}^\mathrm{AA}$. The population of all orthogonal states is negligible.
    (c)~Transfer fidelity as a function of the cooperativity for a range of parameters $a$, $L$ and $w_0$, including flat mirrors, and the experimentally relevant scenarios discussed in \secref{sec:experiment}. The fidelity agrees well with the analytical predictions shown as dashed lines. We use $\Delta_2 = 500 \gamma_\mathrm{a}$ for all setups, although smaller values would suffice for most points.
 }
\end{figure*}

\subsection{Strong-coupling regime}\label{sec:protocols}

Although the cooperativity is an important figure of merit, it does not capture the relative magnitudes of $g$, $\kappa$, and $\gamma_\mathrm{3D}$, which qualitatively impact the dynamics. The strong-coupling regime, where $g \gg\kappa,\gamma_\mathrm{3D}$, is of particular interest as the target atom and the cavity can coherently exchange excitations. Using the results of the previous section, we find that the strong-coupling conditions are equivalent to $\gamma_0 \gg \gamma_\mathrm{a} \gg (1 - \mathcal{R})^2 \gamma_0$, where we assumed that $\gamma_\mathrm{3D} \approx \gamma_\mathrm{a}$, $w_0 \approx a \approx \lambda_0$, and ignored numerical prefactors. The strong-coupling regime is thus accessible for a sufficiently large reflection coefficient $\mathcal{R}$ by a suitable choice of $\gamma_\mathrm{a}$. By varying $\gamma_\mathrm{a}$ it is also possible to reach different regimes. For instance, the Purcell regime, also known as the bad-cavity regime, requires that $\gamma_\mathrm{a} \ll (1 - \mathcal{R})^2 \gamma_0$.

In practice, it may be challenging to identify a transition that satisfies the requirements on $\gamma_\mathrm{a}$ and that is at the same time resonant with the cavity mode. A common approach to overcome this limitation is to employ a Raman scheme, which enables continuous tuning of the effective transition dipole of the target atom~\cite{sorensen2003measurement,douglas2015quantum,gonzalez2015subwavelength,castells2021atomic}. This scheme requires that each target atom contains a $\Lambda$-system, comprising two long-lived states $\ket{\mathrm{g}_1}$ and $\ket{\mathrm{g}_2}$ which are both connected to an excited state $\ket{\mathrm{e}}$ via a dipole transition [see Fig.~\ref{figProt}(a)]. The cavity couples to the transition $\ket{\mathrm{g}_1} \leftrightarrow \ket{\mathrm{e}}$ with coupling strength $g$. The decay rate of the excited state is given by $\gamma_\mathrm{3D}$, as before. Crucially, we assume that the detuning $\Delta_1$ between this transition and the cavity can be varied, which may be realized in practice using light shifts or a static field. A classical field drives the transition $\ket{\mathrm{g}_2} \leftrightarrow \ket{\mathrm{e}}$ with Rabi frequency $\Omega$ and detuning $\Delta_2$ to complete the Raman scheme.

This arrangement leads to a cavity-mediated two-photon transition between $\ket{\mathrm{g}_1}$ and $\ket{\mathrm{g}_2}$, which is resonant when $\Delta_1 = \Delta_2$, ignoring small corrections due to Stark shifts. As shown in App.~\ref{ap:raman}, the excited state can be adiabatically eliminated if $g$ and $\Omega$ are much smaller than the detunings. The two ground states then form an effective two-level system whose transition dipole is reduced compared to the dipole moment of the $\ket{\mathrm{g}_1} \leftrightarrow \ket{\mathrm{e}}$ transition by a factor $\epsilon \approx \Omega / \Delta_1$, where we assumed that $|\Delta_1| \gg \gamma_\mathrm{3D}$. The effective coupling strength and decay rate are thus given by $g_\mathrm{eff} = \epsilon g$ and $\gamma_\mathrm{eff} = \epsilon^2 \gamma_\mathrm{3D}$. Since $g_\mathrm{eff}^2 / \gamma_\mathrm{eff} = g^2 / \gamma_\mathrm{3D}$, the effective cooperativity for the Raman transition $\ket{\mathrm{g}_1} \leftrightarrow \ket{\mathrm{g}_2}$ is the same as for the transition to which the cavity couples. We show in App.~\ref{ap:raman} that this conclusion in fact relies on the stronger condition $|\Delta_{1}| \gg \sqrt{C} \gamma_\mathrm{3D}$, which is needed to suppress additional decoherence caused by the Raman scheme. Provided that this condition is met, we can tune the ratio $g_\mathrm{eff} / \gamma_\mathrm{eff}$ by adjusting $\epsilon$ while keeping the cooperativity constant. The scheme thus enables access to the strong-coupling regime without having to tune the value of $\gamma_\mathrm{a}$.

To demonstrate the effectiveness of this approach, we consider vacuum Rabi oscillations of a single atom placed at the center of an array cavity. We choose identical decay rates for the excited states of the array atoms and the target atom, i.e., $\gamma_0 = \gamma_\mathrm{a}$. This corresponds to the practically relevant scenario where all atoms are identical. The fidelity of vacuum Rabi oscillations is maximized when the effective decay rate of the target atom equals the decay rate of the cavity~\cite{sorensen2003measurement}. This can be achieved using the Raman scheme by setting $\Omega / \Delta_1 = \sqrt{\kappa / \gamma_\mathrm{3D}}$.

In \figref{figProt}(b,left), we show the oscillations for an array cavity with cooperativity $C = 4.3\times 10^4$. We highlight that we did not perform an approximate adiabatic elimination but instead included the full $\Lambda$-system describing the target atom in the calculation. The target atom is prepared in the state $\ket{\mathrm{g}_2}$ at time $t = 0$ while the array atoms all start in the ground state. The excitation is transferred to the cavity mode after the time $t = \pi / 2 g_\mathrm{eff}$. The vacuum Rabi oscillations, however, decay with an envelope $\exp(-\kappa t)$, which leads to the transfer fidelity $F = \exp(- \pi / \sqrt{C})$, where we used the fact that $\kappa$ is equal to the effective free-space decay rate of the target atom.

The Raman scheme can be readily extended to multiple target atoms. In \figref{figProt}(b,right), we plot the excitation probabilities of the $\ket{\mathrm{g}_2}$ state of two target atoms placed at adjacent antinodes close to the center of the array cavity. One of the target atoms is initialized in the state $\ket{\mathrm{g}_2}$, whereas the other target atom is prepared in $\ket{\mathrm{g}_1}$ and the array atoms start in the ground state. The excitation is exchanged between the two atoms after a time $t = \pi / \sqrt{2} g_\mathrm{eff}$. This population exchange can be viewed as the consequence of a complete vacuum Rabi oscillation of the bright state $( \ket{\mathrm{g}_2} \ket{\mathrm{g}_1} + \ket{\mathrm{g}_1} \ket{\mathrm{g}_2} ) / \sqrt{2}$, which couples to the cavity with enhanced Rabi frequency $\sqrt{2} g_\mathrm{eff}$. The dark mode $( \ket{\mathrm{g}_2} \ket{\mathrm{g}_1} - \ket{\mathrm{g}_1} \ket{\mathrm{g}_2} ) / \sqrt{2}$ is decoupled from the cavity and evolves trivially. As above, the oscillations decay at $\exp(-\kappa t)$, which results in the fidelity $F = \exp(- \pi \sqrt{2 / C})$ for the population transfer from one target atom to the other.

We repeated the above computations for different configurations of the array cavities. The resulting fidelities for population exchange is plotted in \figref{figProt}(c) as a function of the cooperativity, which we compute independently for each array cavity. The excellent agreement of the fidelities from the dynamics with the theoretical prediction based on the cooperativity confirms that array cavities can be accurately described by conventional cavity QED parameters. We highlight that our setup could potentially also realize protocols with a more favorable dependence of the fidelity on the cooperativity~\cite{kastoryano2011dissipative}.

\section{Practical considerations}\label{sec:experiment}

\subsection{Cavity transmission}\label{sec:transmission}

\begin{figure}[tb]
	\includegraphics[width=0.99\linewidth]{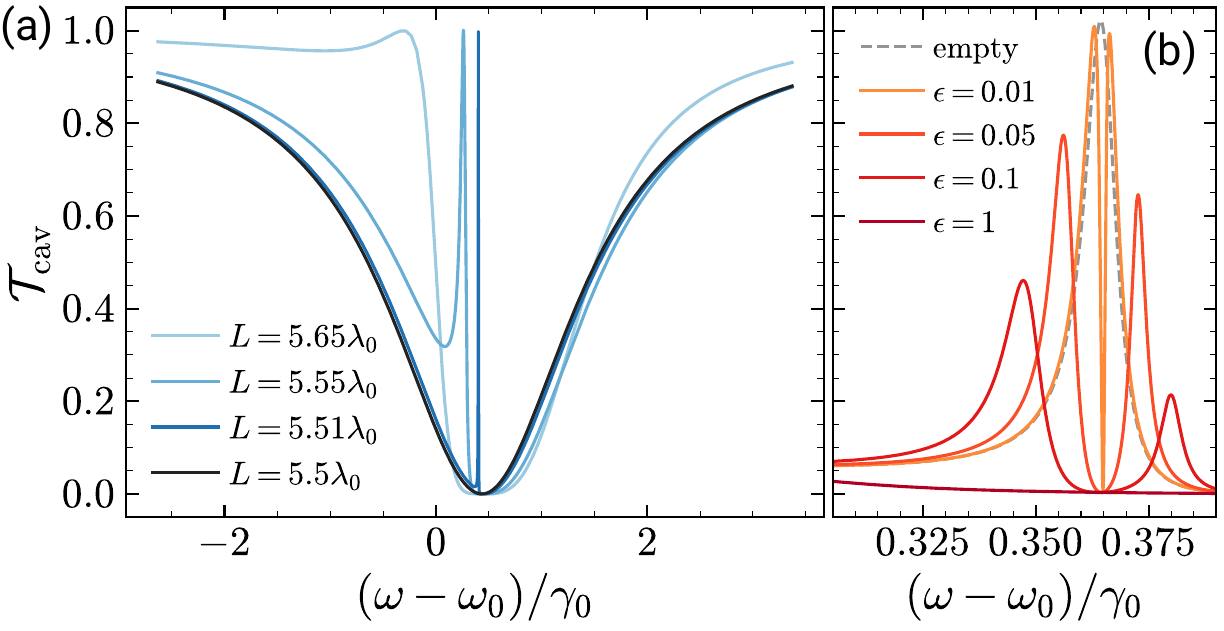}
	\caption{\label{figTransmission} Transmission spectrum of array cavities probed with a Gaussian beam whose wavefront matches the curvature of the arrays. (a)~Transmission through an empty cavity with $a=0.47\lambda_0$, $w_0=4\lambda_0$, $N\times N =40\times 40$, and varying values of $L$ (see legend).
    (b)~Transmission through a cavity with one target atom at its center. The cavity parameters are $a=0.47\lambda_0$, $w_0=1.8\lambda_0$, $L=2.52\lambda_0$, and $N\times N =20\times 20$. We employ the Raman scheme described in \secref{sec:protocols} to vary the linewidth of the target atom as $\gamma_\mathrm{eff}=\epsilon^2\gamma_\mathrm{a}$, which enables us to smoothly interpolate between different cavity regimes. The two-photon resonance condition is always satisfied. The average height of the transmission peaks is approximately $\kappa^2/(\kappa+\gamma_\mathrm{eff})^2$ in the strong coupling regime~\cite{murr2003suppression}.	
 }
\end{figure}

We have focused so far on the internal dynamics of the array cavity and the target atom. In this section, we extend our analysis to include the incident and scattered fields, which are of great relevance for many applications in, e.g., quantum communication. In Fig.~\ref{figTransmission}(a), we show the transmission spectrum of a Gaussian probe beam impinging on an array cavity for different values of the cavity length, $L$. The waist of the Gaussian beam is matched to the curvature of the arrays. We evaluate the intensity transmission coefficient, $\mathcal{T}_\mathrm{cav}$, and reflection coefficient, $\mathcal{R}_\mathrm{cav}$, by projecting the scattered field onto the same Gaussian mode as detailed in Appendix~\ref{ap:transmission}. The cavity transmission shows a Lorentzian dip of width $\sim \Gamma_0$, which arises from the reflection resonance of the individual mirrors. In addition, there is a cavity-like resonance, whose width and location depends sensitively on the cavity length.

As discussed in~\cite{pedersen2023quantum,bassler2024metasurface}, the transmission spectrum is accurately described by the Fabry-P\'erot formula $\mathcal{T}_\mathrm{cav}(\omega) = \left| t(\omega)^2 / [1 - r(\omega)^2 e^{2 i \omega L / c}] \right|^2$, where $r(\omega)$ and $t(\omega)$ are the amplitude reflection and transmission coefficients of a single array mirror. The frequency of the cavity resonance is determined by the interplay of the propagation phase, which depends on the cavity length, and the frequency-dependent phase of $r(\omega)$. When $L$ is an integer multiple of $\lambda_0 / 2$, the cavity resonance occurs at the center of the resonance of a single mirror. There is no discernible transmission peak in Fig.~\ref{figTransmission}(a) for this case because the transmission $t(\omega)$ is small compared to scattering losses (see below). As we vary $L$, the cavity resonance moves away from the resonance of the mirror and consequently the relevant value of $t(\omega)$ increases. This causes the cavity resonance to broaden and to become clearly visible in transmission. We note that the transmission spectrum can alternatively be viewed as the result of interference between bright and dark modes corresponding to in-phase and out-of-phase excitation of the two mirrors~\cite{guimond2019subradiant,pedersen2023nonlinear}.

\begin{figure*}[tb]
	\includegraphics[width=0.99\linewidth]{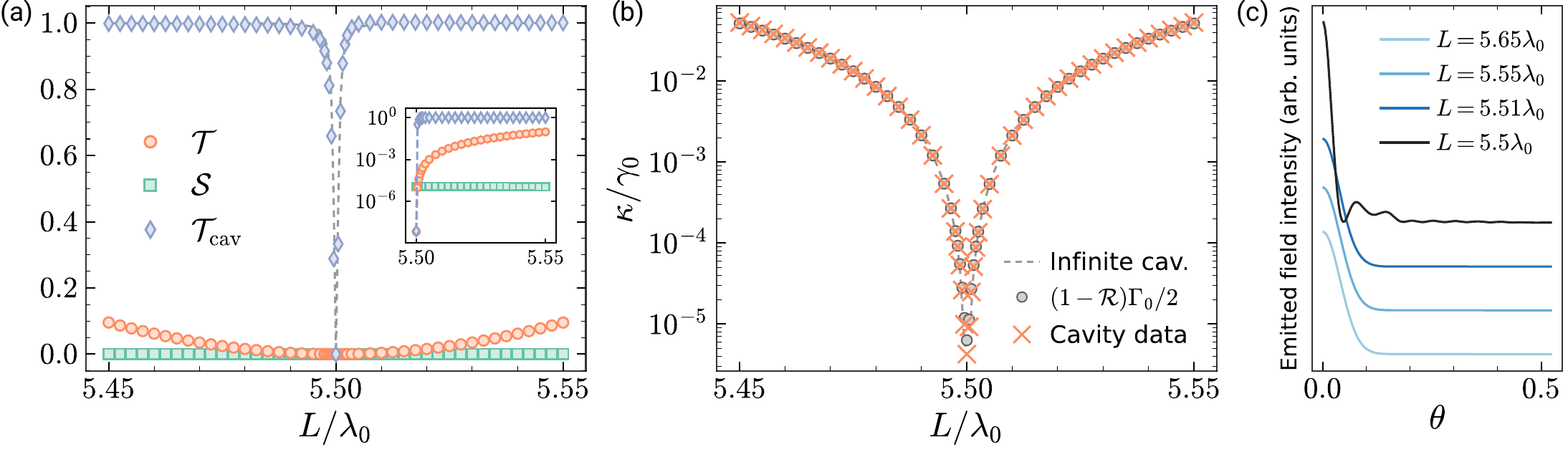}
	\caption{\label{figMaxTransmission} Dependence of the cavity properties on the separation between the array mirrors. We show the results for arrays with the same parameters as in \figref{figTransmission}(a). (a)~Transmission, $\mathcal{T}$, and scattering loss, $\mathcal{S}$, of a single mirror at a frequency shifted by $\tan(k_0 L)\Gamma_0/2$ from the mirror resonance. The shift corresponds to the location of the cavity transmission peak. We also show the height of the cavity transmission peak (blue diamonds) and compare it to the expectation for a conventional cavity, $\mathcal{T}_\mathrm{cav} = \mathcal{T}/(\mathcal{T}+\mathcal{S})$ (dashed curve). The inset shows the same data in semi-logarithmic scale.
    (b)~Comparison between the cavity decay rate computed as in \secref{sec:kappa} (orange crosses) and the prediction of \eqref{eq:kappa} using the reflection coefficient of a single mirror at the cavity resonance frequency (gray circles). We also show the analytical prediction $\kappa \approx (k_0\delta L)^2\Gamma_0/2$ (dashed curve) derived for infinite arrays whose length differs by a small amount from an integer multiple of $\lambda_0 / 2$.
    (c)~Far-field electric field intensity emitted by the cavity eigenstate (see Appendix~\ref{ap:transmission}) as a function of the polar angle, $\theta$, and averaged over the azimuthal angle. The data have been rescaled and shifted for illustrative purposes. The emission pattern deviates significantly from a Gaussian at $L = 5.5 \lambda_0$, where the cavity resonance coincides with the resonance of the array mirrors and transmission through the mirrors is strongly suppressed.
	}
\end{figure*}

The presence of a target atom can strongly modify the transmission spectrum. In Fig.~\ref{figTransmission}(b), we show the transmission of an array cavity with a target atom placed at the center. We take $\gamma_\mathrm{a}=\gamma_0$ and tune the linewidth of the target atom using a Raman transition as described in \secref{sec:protocols}. The single-photon detuning is again assumed to be large, and the cavity is resonant with the two-photon transition. For $\epsilon \approx 1$, we have $\epsilon^2 \gamma_\mathrm{3D} > \epsilon g > \kappa$. In this regime, the target atom has a large absorption, resulting in the suppression of the cavity transmission. For smaller values of $\epsilon$, we access the strong coupling regime, where two peaks separated by the vacuum Rabi splitting $2 \epsilon g$ can be resolved. In all regimes, the presence of the target atom suppresses the transmission at the cavity resonance by a factor $1 / (C + 1)^2$, which can be used to realize quantum gates between photons and the target atom~\cite{thompson2013Coupling}.

The maximum transmission through the cavity is limited by scattering into modes with little spatial structure. For a single mirror, we define $\mathcal{R}$ and $\mathcal{T}$ as the intensity reflection and transmission coefficients for a single collimated mode, which can be interfaced with conventional far-field optics. We always project onto the Gaussian beam determined by the length of the cavity and the curvature of the array mirror. The scattering loss coefficient is given by ${\mathcal{S}=1-\mathcal{R}-\mathcal{T}}$. To compute the transmission through the cavity, these coefficients must be evaluated at the cavity resonance.

As shown in \figref{figMaxTransmission}(a), the scattering loss $\mathcal{S}$ is approximately independent of the resonance frequency determined by the cavity length. By contrast, the transmission coefficient satisfies $\mathcal{T} \approx \mathcal{T}_0 + (k_0 \delta L)^2$, where $\delta L$ is the difference of the cavity length from the nearest integer multiple of $\lambda_0 / 2$, assuming $k_0 \delta L \ll 1$. This follows from the fact that $\mathcal{T}$ is approximately a Lorentzian function of frequency with width $\Gamma_0$ and that the cavity resonance is shifted from the resonance of the mirror by $\tan (k_0 \delta L) \Gamma_0 / 2$~\cite{pedersen2023nonlinear}. Just as for a conventional cavity, we expect the maximum transmission at the cavity resonance to be given by $\mathcal{T}_\mathrm{cav} = \mathcal{T} / (\mathcal{T} + \mathcal{S})$. This expression, indicated by the dashed line in \figref{figMaxTransmission}(a), is indeed in excellent agreement with the data for array cavities of differenth lengths. Cavity transmission is strongly suppressed at $\delta L = 0$ because $\mathcal{S} \gg \mathcal{T}_0$, i.e., the scattering losses are much greater than transmission through the array mirror.

For many purposes, it is useful to distinguish between cavity decay due to transmission and due to loss. Following \secref{sec:kappa}, the corresponding cavity decay rates are given by $\kappa_\mathrm{out} = \mathcal{T} \, \Gamma_0 / 2$ and $\kappa_\mathrm{loss} = \mathcal{S} \, \Gamma_0 / 2$. From their sum, we recover the total cavity decay rate in agreement with \eqref{eq:kappa}. The nonzero value of $\kappa_\mathrm{loss}$ explains the saturation in the scaling of the cooperativity at small $\mathcal{T}$ in \figref{figIntro}(c). \Figref{figMaxTransmission}(b) shows that this approach of computing the cavity decay rate using the reflection coefficient of a single mirror agrees well with the value of $\kappa$ obtained from the imaginary part of the eigenvalue of $\vec{H}^\mathrm{AA}$ corresponding to the cavity eigenmode. The only significant deviation occurs close to $\delta L = 0$, where the values of $\kappa$ obtained from the mirror properties are an overestimate. We attribute this discrepancy to the profile of the cavity mode, which notably differs at those cavity lengths from the Gaussian beam used to probe the reflectivity of the mirrors. This is evident in Fig.~\ref{figMaxTransmission}(c), where we show the far-field emission from the cavity eigenstate computed using \eqref{eq:IO}. The emission profile matches the Gaussian mode set by the cavity length and the curvature of the mirrors for sufficiently large $\delta L$. At $\delta L = 0$, however, the emission pattern is distinctly non-Gaussian. Nevertheless, the emitted field remains collimated within a small solid angle, thereby allowing for efficient detection.

\begin{figure}[tb]
	\includegraphics[width=0.99\linewidth]{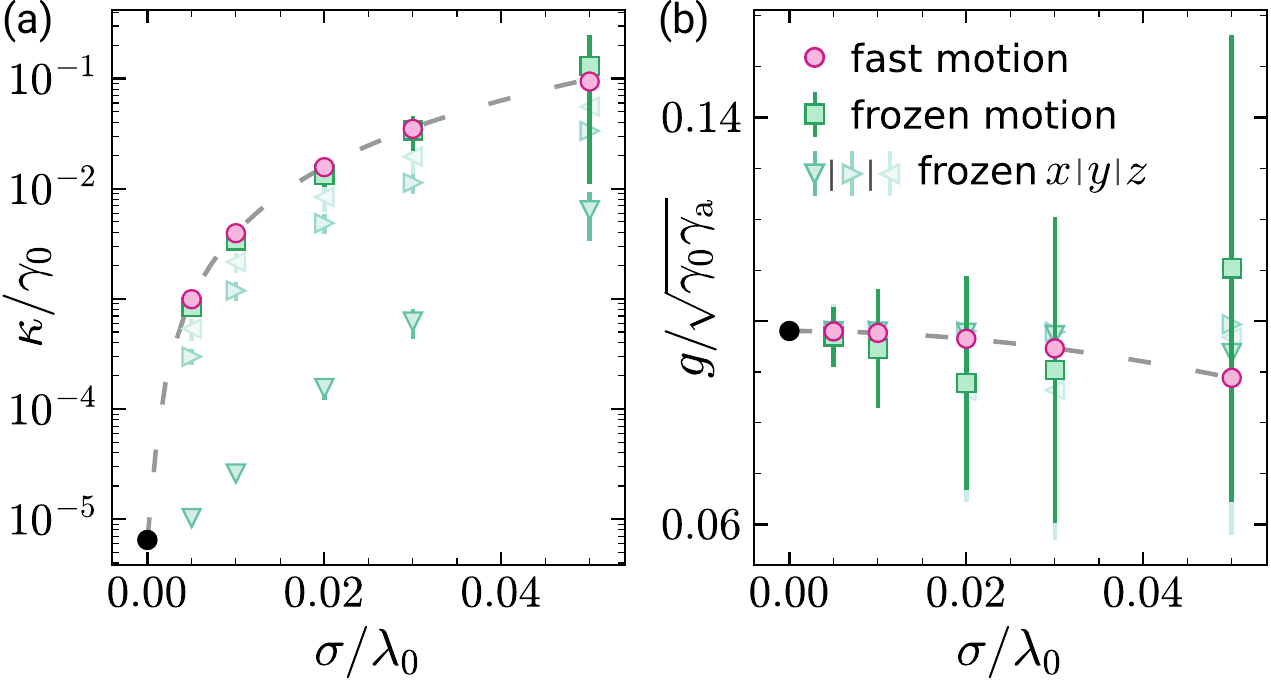}
		\caption{\label{figMotion} Effect of motion of the array atoms. (a)~Cavity decay rate and (b)~coupling strength as a function of the root-mean-square displacement $\sigma$. In both plots, we consider fast and frozen motion for an array cavity with $a=0.47\lambda_0$, $w_0=3\lambda_0$, $L=1.5\lambda_0$, and $N\times N=30\times 30$. In the frozen motion regime, we average over 200 disorder realizations and also show the results when the atoms are only free to move along the $x$ axis (direction of the transition dipole), $y$ axis, or $z$ axis (cavity axis). The dashed lines represent the prediction from the fast-motion Hamiltonian in the Lamb-Dicke limit. The black dots indicate the results without motion.
 }
\end{figure}

\subsection{Motion and disorder}\label{sec:motion}

So far, we have considered the ideal case of point dipoles in a perfectly ordered lattice. For a realistic prediction, however, we need to include position fluctuations, which persist even in the motional ground state of a trapping potential. Motion deteriorates the properties of the array mirrors, as experimentally observed in Ref.~\cite{rui2020subradiant}, because the near-perfect reflection relies on the collective interplay of dipole--dipole interactions, which depend strongly on the relative positions of the atoms. Although a full quantum treatment of atomic motion is computationally prohibitive due to the many atoms involved, we can estimate the impact of motion on the cavity parameters by considering two extreme regimes distinguished by the speed of the motion compared to the characteristic time scale of the internal dynamics~\cite{porras2008collective}.

In the \emph{frozen-motion regime}, we assume that the positions of the atoms fluctuate much more slowly than the slowest internal dynamics. Assuming that the atoms are located in harmonic traps with trapping frequency $\nu_\mathrm{T}$, this corresponds to the condition $\nu_\mathrm{T} \ll \mathrm{min} \{g_\mathrm, \kappa, \gamma_\mathrm{3D}\}$~\footnote{In the case of the Raman dressing scheme, $g$ and $\gamma_\mathrm{3D}$ should be replaced by the effective quantities rescaled by $\epsilon$ and $\epsilon^2$, respectively.}. Each atom may then be viewed as fixed at a certain position during the time evolution. We obtain instances of such disordered realizations by sampling the position of the $i^\mathrm{th}$ atom from the distribution $p_i(\vec{r})=\exp[-(\vec{r}-\vec{r}_i)^2/2\sigma^2]/(\sqrt{2\pi}\sigma)^3$, where $\vec{r}_i$ is the ideal position. For simplicity, we take the standard deviation $\sigma$ to be the same in all three spatial directions. The expectation value of physical observables, such as $g$ and $\kappa$, are computed by averaging over many disordered realizations. This approach has been shown to be exact in the limit of unsaturated atoms with infinite mass~\cite{lee2016stochastic}.

In the \emph{fast-motion regime}, we consider the position fluctuations to be much faster than the internal dynamics, $\nu_\mathrm{T}\gg\gamma_0$. As discussed in previous works~\cite{guimond2019subradiant,rusconi2021exploiting}, the motional degrees of freedom can be adiabatically eliminated in this limit, which corresponds to averaging the dipole-dipole interaction over the position distribution of the atoms. Using the same position distributions $p_i(\vec{r})$ as for the frozen case, we show in Appendix~\ref{ap:fastmotion} that the resulting Hamiltonian is given by
\begin{align}\label{eq:fastmotion}
    \hat{H}_\mathrm{fast} = &\sum_i H_{ii}\,\hat{\sigma}^+_i\hat{\sigma}^-_i - e^{-k_0^2\sigma^2/2} \sum_i \left( {\Omega}_i\hat{\sigma}_i^+ +\mathrm{h.c.} \right) \nonumber \\
    &+ e^{-k_0^2\sigma^2}\sum_{i\neq j}\left(\Delta_{ij}-\frac{i}{2}\Gamma_{ij}\right)\hat{\sigma}^+_i\hat{\sigma}^-_j ,
\end{align}
where $H_{ii}$ denotes the original coefficient of the $\hat{\sigma}^+_i\hat{\sigma}^-_i$ term in \eqref{eq:drivenHam}. The above expression is valid in the Lamb-Dicke limit, where $\eta=k_0\sigma\ll1$.

In Fig.~\ref{figMotion}, we show $\kappa$ and $g$ as a function of $\sigma$ for the two regimes. Motion causes scattering loss, resulting in an additional contribution to the cavity decay rate that is approximately given by $\kappa_\mathrm{loss}^\mathrm{mot} = \eta^2\gamma_0$ in both regimes [dashed curve in \figref{figMotion}(a)]. For fast motion, this decay rate results from imperfect cancellation of the individual free-space decay rate $\gamma_0$ due to the suppression of the dipole--dipole interaction by the factor $e^{-\eta^2}$ in \eqref{eq:fastmotion}. We also show the results for frozen motion when we only add disorder in one of the three spatial dimensions. The impact of motion on $\kappa$ is the largest in the $z$ direction, orthogonal to the arrays, and smallest in the $x$ direction. We attribute the latter to the fact that the atoms are polarized along the $x$ axis such that the interaction in this direction is much weaker due to the dipole emission pattern. We have verified this claim by repeating the computation for atoms with a circularly polarized transition, in which case frozen disorder has identical effects in both in-plane directions. The coupling strength $g$ is affected less significantly by the motion as it is merely rescaled by $e^{-\eta^2}$ [dashed curve in \figref{figMotion}(b)]. For fast motion, this modification can again be understood in terms of the modified dipole-dipole interaction in \eqref{eq:fastmotion}.

We remark that the quantitative agreement of frozen and fast motion does not hold universally. We numerically observe that the loss in the two regimes is similar when the lattice constant satisfies $a \gtrsim 0.4\lambda_0$. When $a \lesssim 0.4 \lambda_0$, frozen motion affects the cavity lifetime more severely whereas the impact of fast motion is independent of $a$. The enhanced sensitivity to frozen motion is caused by a degeneracy of the cavity mode with other subradiant eigenstates of the array. The static position disorder, which breaks the order of the array, induces detrimental hybridization of these modes. Small lattice constants should therefore be avoided in practice unless the (quasi) static positions of the atoms can be controlled to a high degree. Motional sidebands, which are relevant in intermediate regimes between frozen and fast motion, can also give rise to similar hybridization of the cavity mode with states detuned by an energy equal to the trap frequency.

The above results indicate that losses due to motion are in many cases the dominant contribution to $\kappa$ and therefore have a significant impact on the cooperativity. In this case, the highest cooperativities are achieved by cavities that maximize the coupling strength $g$ while the value of the intrinsic cavity linewidth, previously discussed in \secref{sec:kappa}, is unimportant.  To estimate experimentally achievable values of the cooperativity, we consider an optical lattice as in \cite{rui2020subradiant,srakaew2023subwavelength}, where three pairs of counterpropagating laser beams generate a three-dimensional, periodic trapping potential. For simplicity, we assume that the trapping potential is isotropic with trap depth $V_0$. Assuming deep traps, such that tunneling is weak, the trapping potential around the minima can be approximated by harmonic oscillators, with trap frequency $\nu_\mathrm{T} = 2 \sqrt{V_0 E_\mathrm{r}} / \hbar$. Here, $E_\mathrm{r}=h^2/(8ma^2)$ is the recoil energy of an atom, which are assumed to all have equal mass $m$. The Wannier functions of the atoms are approximately Gaussian, characterized by the oscillator length $l = \sqrt{\hbar / 2 m \nu_\mathrm{T}}$.

We expect the lowest losses when all atoms are in their vibrational ground states such that $\sigma = l$. Following the preceeding discussion, we estimate the contribution to the cavity decay rate due to motion as
\begin{equation}
    \kappa_\mathrm{loss}^\mathrm{mot} \approx \eta^2\gamma_0  = 2 \left( \frac{a}{\lambda_0} \right)^2 \sqrt{\frac{E_\mathrm{r}}{V_0}}\;\gamma_0\,.
\end{equation}
By assuming that this rate is the dominant source of cavity decay, we obtain the estimate
\begin{equation}\label{eq:coopmot}
    C \approx \frac{4 g^2}{\kappa_\mathrm{loss}^\mathrm{mot} \gamma_\mathrm{3D}} \approx \frac{9}{8\pi^3} \left( \frac{\lambda_0}{w_0} \right)^2 \left( \frac{\lambda_0}{a} \right)^4 \sqrt{\frac{V_0}{E_\mathrm{r}}}\;\frac{\gamma_\mathrm{a}}{\gamma_\mathrm{3D}}
\end{equation}
for the cooperativity. Here, we used the value of $g$ as estimated without motion.

Our results indicate that the cooperativity can be maximized by minimizing the beam waist and lattice spacing while maximizing the trap depth. We note, however, that the validity of the expression is limited to regimes without motion-induced hybridization as discussed above, and that heating places a practical constraint on $V_0$ if the excited state is anti-trapped~\cite{rui2020subradiant,karanikolaou2024near}. We circumvent the latter concern by assuming that the atoms are trapped at a magic wavelength. For the $D_2$ transition in $^{87}\mathrm{Rb}$ ($\lambda_0 \approx 780~\mathrm{nm}$) such a magic wavelength exists at $\lambda_\mathrm{T} \approx 740~\mathrm{nm}$~\cite{arora2007magic}, this corresponds to $a = \lambda_\mathrm{T} / 2 \approx 0.47 \lambda_0$. Choosing a deep, yet realistic, lattice depth of $V_0=2000E_\mathrm{r}$ and a nearly diffraction limited beam waist of $w_0 = 2 \lambda_0$, \eqref{eq:coopmot} yields a cooperativity of $\gamma_\mathrm{3D}C/\gamma_\mathrm{a} \approx 8.3$, in agreement with \figref{figIntro}(c) considering that $\gamma_\mathrm{3D}\simeq0.5\gamma_\mathrm{a}$ for $L\simeq 1.5\lambda_0$. We remark that these parameters correspond to an intermediate regime between frozen and fast motion since the trap frequency $\nu_\mathrm{T} \approx 2 \pi \times 0.4~\mathrm{MHz}$ is much smaller than the $D_2$ linewidth $\gamma_0 \approx 2 \pi \times 6~\mathrm{MHz}$ but greater than the cavity linewidth $\kappa_\mathrm{loss}^\mathrm{mot} \approx 2 \pi \times 0.06~\mathrm{MHz}$.

\begin{figure*}[tb]
	\includegraphics[width=\linewidth]{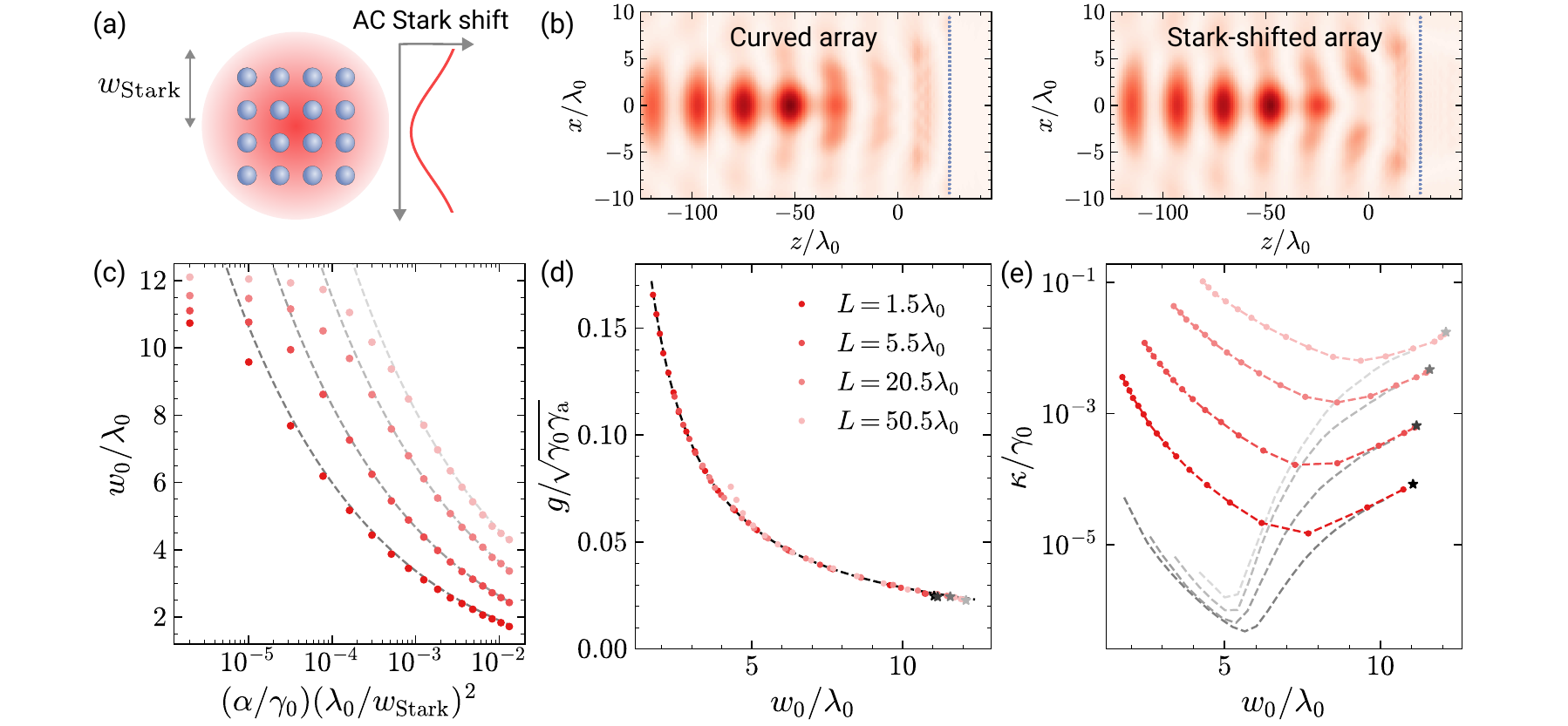}
	\caption{\label{figStark} Effective curving of the mirrors with a position-dependent AC Stark shift.
    (a)~The profile of a wide Gaussian beam produces a local detuning on the lattice atoms [see \eqref{eq:shiftStark}], which imparts a position-dependent phase on the reflected light that mimics a curved mirror.
    (b)~Electric field intensity for a plane wave incident from the left on a curved array (left panel) and a flat array with a position-dependent Stark shift (right panel). The array parameters are $a=0.47\lambda_0$ and $N\times N=40\times40$ in both cases and the frequency of the incident light is chosen to maximize reflection. The Stark shift on the flat array, quantified by $\alpha/w_\mathrm{Stark}^2=7\times 10^{-3}\gamma_0/\lambda_0^2$, has been adjusted to match the radius of curvature $R=270\lambda_0$ of the curved array. The maximum of the color scale for the Stark-shifted array is approximately $30 \%$ lower than that of the curved array.
    (c)~Dependence of the beam waist, $w_0$, on the strength of the Stark shift for a cavity with $a=0.47\lambda_0$, $N\times N=60\times60$. We obtain $w_0$ by fitting a Gaussian to the wavefunction of the cavity eigenstate on the mirror. We fix $w_\mathrm{Stark}=500\lambda_0$ although the exact value is unimportant. Deviations from the prediction [dashed curves, \eqref{eq:effwaist}] are due to the finite size of the arrays.
    (d)~Dependence of the coupling strength, $g$, on the fitted value of $w_0$ for the same cavities as in (c). The stars are the results for flat mirrors without the AC Stark shift. As for curved mirrors, the coupling strength follows \eqref{eq:g}, represented by the dashed line.
    (e)~By contrast, the cavity decay rate $\kappa$ under the AC Stark differs significantly from value with curved mirrors with the same beam waist (dashed curves). Smaller values of $w_0$, corresponding to larger values of $\alpha$, lead to increased loss.
 }
\end{figure*}

\subsection{Optically induced curvature}\label{sec:stark}

The highest cooperativities computed above were obtained using curved array mirrors. When the atoms are trapped in an optical lattice, the curvature may in principle be created using the optical force of a focused laser or by means of a static field gradient. Owing to the accurate positioning required, this may, however, pose a significant practical challenge. In this section, we propose an alternative to curving the mirrors based on a position-dependent AC Stark shift, as shown in Fig.~\ref{figStark}(a). The AC Stark shift induces a position-dependent phase that mimics the phase dependence of the wavefront of a Gaussian beam as illustrated in Fig.~\ref{figStark}(b).

To analyze this effect, we suppose that the AC Stark shift is caused by a Gaussian beam with width $w_\mathrm{Stark}$ at the location of the mirror. The local detuning is given by
\begin{equation}\label{eq:shiftStark}
    \delta_\mathrm{Stark}(r) = \alpha \left( 1 - e^{- 2 r^2 / w_\mathrm{Stark}^2} \right) \approx \frac{2 \alpha r^2}{w_\mathrm{Stark}^2}  ,
\end{equation}
where the coefficient $\alpha$ is set by the intensity of the Stark shift beam. We added a global offset to zero the shift at the center. The expansion of the exponential to first order is valid provided the width $w_\mathrm{Stark}$ is much larger than the size of the array mirrors. Since the mirror has a Lorentzian response with width $\Gamma_0$, the detuning causes the light in the cavity to acquire a phase shift
\begin{equation}
    \Delta \phi(r) = \frac{2 \delta_\mathrm{Stark}(r)}{\Gamma_0} \approx \frac{4 \alpha r^2}{\Gamma_0 w_\mathrm{Stark}^2}
\end{equation}
upon reflection, where we assumed that the central portion of the incident light is resonant with the mirror.

This phase shift results in an effective curvature of the wavefront. For a cavity with curved mirrors, the corresponding phase shift is given by $\Delta \phi(r) = k_0 r^2 / R(L/2)$. We recall that the curvature of the mirrors is related to the waist by $R(L/2) = (L/2) + (k_0^2 w_0^4 / 2 L) \approx k_0^2 w_0^4 / 2 L$, where we restrict ourselves to the regime $k_0 w_0^2 \gg L$, which applies to the examples presented here. By equating the two expressions for the phase shift, we predict that the AC Stark shift induces an effective curvature that results in a cavity mode with beam waist
\begin{equation}\label{eq:effwaist}
    w_{0} = \left(\frac{L \lambda_0 w_\mathrm{Stark}^2}{4 \pi} \frac{\Gamma_0}{\alpha} \right)^{1/4}\,.
\end{equation}

We verify this prediction by computing the cavity modes of an array cavity as described in \secref{sec:cavitymodes} including the position-dependent detuning. To extract the beam waist $w_0$, we fit a Gaussian to the wavefunction on one of the mirrors. This yields the beam width $w(L/2)$, from which the beam waist can be computed using the relation $w(L/2) = \sqrt{ w_0^2 + (L / k_0 w_0)^2}$. The obtained value is in excellent agreement with \eqref{eq:effwaist} as shown in \figref{figStark}(c). We note that the values saturate at small $\alpha$ due to the finite size of the arrays. We also compute the coupling strength $g$ and the cavity decay rate $\kappa$ using the approaches described in \secref{sec:parameters}. \Figref{figStark}(d) shows that the analytic expression for $g$, \eqref{eq:g}, remains valid when the waist is induced by the AC Stark shift instead of the curvature of the mirrors. However, as shown in \figref{figStark}(e), the Stark shift leads to an enhanced cavity decay rate compared to curved mirrors corresponding to the same beam waist. We attribute this to the fact that the detuned atoms modify the collective resonance of the array, reducing its reflectivity. The enhanced decay rate may not be detrimental in practice, where we expect the main contribution to $\kappa$ to be caused by the motion of the atoms.

We note that the cavity length that yields the minimum values of $\kappa$ deviates slightly from integer multiples of $\lambda_0/2$ due to the Gouy phase of the confined mode. We have accounted for this effect in Figs.~\ref{figStark}(c--e) by adjusting $L$ in the proximity of the stated values such that $\kappa$ is at a local minimum. The observed correction to $L$ is proportional to $L / (k_0 w_0)^2$, in accordance with the expectation from the Gouy phase.

\section{Discussion and outlook\label{sec:conclusion}}

\begin{figure}[tb]
	\includegraphics[width=\linewidth]{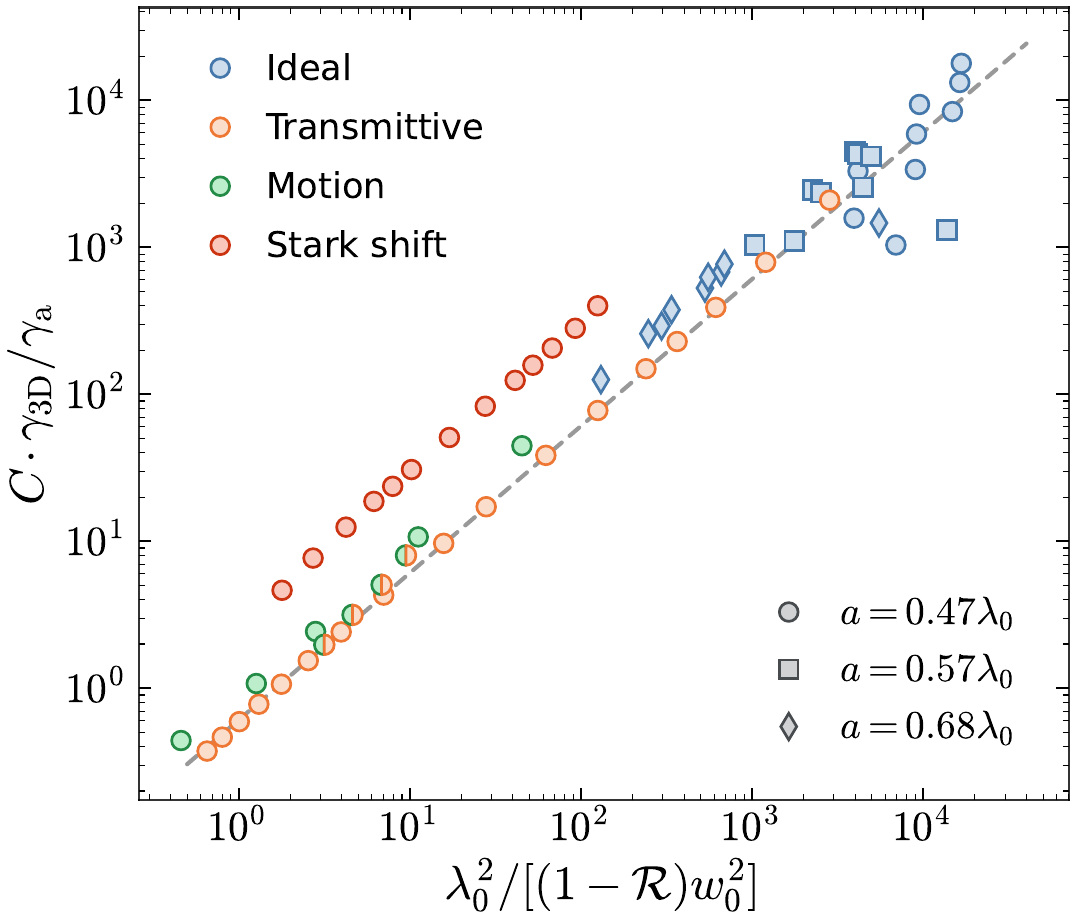}
	\caption{\label{figAtomConv}
            Cooperativity, $C$, as a function of $\lambda_0^2/ [(1-\mathcal{R})w_0^2]$, where $\mathcal{R}$ is the reflection coefficient of a single mirror at the resonance frequency of the cavity. We provide the range of parameters of the cavities in Appendix~\ref{ap:fig10}. The points labeled ``ideal'' correspond to the cavities considered in \secref{sec:atomcav}, where the atoms are fixed and $L$ is an integer multiple of $\lambda_0 / 2$. The labels ``transmittive'', ``motion'', and ``Stark shift'' refer to the practical scenarios discussed in \secref{sec:transmission}, \ref{sec:motion}, and \ref{sec:stark}, respectively. We scale the cooperativity by $\gamma_\mathrm{3D}/\gamma_\mathrm{a}$ to highlight the dependence on $\mathcal{R}$ and $w_0$. All data points agree well with the theoretical prediction, \eqref{eq:c} (dashed line), which also holds for conventional cavities.
	}
\end{figure}

Our work establishes atom arrays as a promising platform for cavity QED. We show that atom array cavities can be described by the same parameters as conventional cavity QED setups. The coupling strength $g$ and the cavity decay rate $\kappa$, however, are reduced compared to conventional setups due to the time delay that light experiences when reflected by a narrow-band mirror. The reduction in these two parameters cancels when computing the cooperativity $C = 4 g^2 / \kappa \gamma_\mathrm{3D}$. Hence, array cavities and conventional cavities with equivalent mirrors in terms of curvature and reflection result in the same cooperativity. We highlight that in the case of array cavities, the cavity length determines the detuning of the cavity mode from the resonance of the array mirrors. The reflection coefficient of the mirror should therefore be viewed as a function of the cavity length.

\Figref{figAtomConv} summarizes these findings by showing the cooperativity as a function of $\lambda_0^2 / [ (1 - \mathcal{R}) w_0^2 ]$ for a variety of configurations, including the limitations considered in \secref{sec:experiment}. We show $\gamma_\mathrm{3D} C / \gamma_\mathrm{a}$ instead of $C$ to simplify comparison as the value of $\gamma_\mathrm{3D}$, which ranges from $0.5 \, \gamma_\mathrm{a}$ to $\gamma_\mathrm{a}$ for the array cavities displayed in the figure, would lead to increased scatter. The computed cooperativities are all close to the analytical prediction [\eqref{eq:c}, dashed line] based on the beam waist and the separately obtained reflection coefficient of a single array mirror. We observe the most significant deviation for array cavities where the beam waist is defined using an AC Stark shift (red circles). The discrepancy is explained by the fact that the AC Stark shift results in a weakly non-Gaussian profile of the cavity mode. This in turn leads to a systematic underestimate of the reflection coefficient because $\mathcal{R}$ is computed by projecting onto a Gaussian beam. The same reasoning applies to cavities with minimal transmission, where $L$ is an integer multiple of $\lambda_0/2$ [c.f.~\figref{figMaxTransmission}(c)]. For the points below the dashed line, we hypothesize that their cooperativity is reduced due to coupling between different cavity modes, which results in loss not captured by the reflection coefficient of a single mirror.

Our work focuses on maximizing the cooperativity as a key figure of merit for applications in quantum communication protocols. We highlight that simpler configurations may be sufficient to demonstrate the experimental viability of the proposed platform in the near term. For instance, our numerical results indicate that small flat arrays, where the transverse confinement of the cavity mode is due to boundary effects, can achieve cooperativities exceeding unity. Concretely, using $a=0.47\lambda_0$, $L=1.5\lambda_0$, $N\times N=10\times10$, and $V_0=2000E_\mathrm{r}$, we obtain $C\simeq 4$. The cooperativity can be subsequently improved by adding a position-dependent AC Stark shift. This scheme may be of interest in its own right and could, for instance, be used to create metalenses from flat atom arrays. 

There are several avenues to improving the properties of array cavities. The cooperativity can be increased by a tighter confinement with higher laser intensities, assuming trapping at a magic wavelength. To reduce the sensitivity to motion and disorder, one could optimize the individual positions of the atoms beyond the simple curved arrays considered here. Scattering may be suppressed by closing the cavity with a cylindrical array, albeit at the cost of increased experimental complexity. We emphasize that our proposal is not limited to ultracold atoms in optical lattices but applies to arrays of dipoles in general. It may be fruitful in this context to explore realizations based on optical tweezer arrays, which allow for versatile geometries~\cite{kaufman2021quantum}. The limitations due to motion could be overcome by using solid-state emitters, although such systems often suffer from detrimental inhomogeneous broadening~\cite{schroeder2017scalable}. Finally, similar phenomena may be explored in two-dimensional semiconductors, where delocalized excitons take the role of the discrete emitters~\cite{back2018realization,scuri2018large,wild2018quantum}.

Our results establish a theoretical foundation for the description of array cavities using the language of conventional cavity QED. It will be exciting to move beyond this framework to explore phenomena that take advantage of features of atom arrays not accessible with conventional mirrors. A particularly intriguing direction is to explore the dynamics of two or more excitations since atom-array cavities have been shown to exhibit quantum nonlinearities at the single-photon level~\cite{pedersen2023quantum,robicheaux2023Intensity}. In combination with the ability to dynamically control the arrays, this approach may enable on-demand generation of complex quantum states of light.

\begin{acknowledgments}
    We thank Ana Asenjo Garc\'ia, Darrick Chang, Pau Farrera, Daniel Malz, Simon Panyella Pederson, Cosimo Rusconi, and Peter Zoller for insightful discussions. We thank Daniel Adler, Suchita Agrawal, Kritsana Srakaew, Pascal Weckesser, David Wei, and Johannes Zeiher for sharing their expertise concerning the experimental realization of atom arrays. We also thank Johannes Zeiher for valuable feedback on the manuscript. This research is part of the Munich Quantum Valley (MQV), which is supported by the Bavarian State Government with funds from the High-tech Agenda Bayern Plus. We acknowledge funding from the Munich Center for Quantum Science and Technology (MCQST), funded by the Deutsche Forschungsgemeinschaft (DFG) under Germany’s Excellence Strategy (EXC2111-390814868). D.~S.~W. received funding from the European Union’s Horizon 2020 research and innovation programme under the Marie Skłodowska-Curie Grant Agreement No.~101023276.
\end{acknowledgments}

\clearpage
\appendix

\section{Effective Hamiltonian}\label{ap:effHam}

The derivation of \eqref{eq:drivenHam} is based on the Born-Markov approximation, which is satisfied self-consistently if the frequency scales that appear in the Hamiltonian are small compared to $\omega_\mathrm{a}$. The approximation is usually excellent for optical transitions, for which $\gamma_i \ll \omega_\mathrm{a}$. In addition, \eqref{eq:drivenHam} neglects the effects of retardation, which requires that $c / \gamma_i$ is much greater than the largest separation between any pair of atoms. This is again an excellent approximation in our case as all length scales are proportional to the wavelength and the typical scale for the linewidth of dipole-allowed optical transitions is MHz.

The photon-mediated terms of the effective Hamiltonian, \eqref{eq:drivenHam}, are proportional to the Green's function $\vec{G}(\vec{r}_i,\vec{r}_j;\omega_0)$, which describes the electric field at position $\vec{r}_i$, generated by a dipole source at position $\vec{r}_j$. In particular, for $i \neq j$,~\cite{asenjo2017exponential}
\begin{equation}
    \Delta_{ij} - \frac{i}{2} \Gamma_{ij} = - \frac{3 \pi c}{\omega_0} \sqrt{\gamma_i \gamma_j} \, \vec{e}_i^* \cdot \vec{G}(\vec{r}_i, \vec{r}_j; \omega_0) \cdot \vec{e}_j ,
    \label{eq:coupling}
\end{equation}
where $\vec{e}_i = \vec{d}_i / |\vec{d}_i|$ are unit vectors corresponding to the polarization of the transition. Throughout this work, the atoms are placed in free space, where
\begin{align}\label{eq:Greensfn}
    \begin{split}
    \vec{G}(\vec{r}_i,\vec{r}_j;\omega_0) = &\frac{e^{ik_0 r}}{4\pi r}\left[\left(1+\frac{ik_0r-1}{k_0^2r^2}\right)\vec{I}\right.\\
    &\left.+\left(-1+\frac{3-3ik_0r}{k_0^2r^2}\right)\frac{\vec{r}\otimes \vec{r}}{r^2}\right]\, .
    \end{split}
\end{align}
Here, $\vec{r} = \vec{r}_i-\vec{r}_j$, $r=|\vec{r}|$, and $\vec{I}$ is the $3 \times 3$ identity matrix. For $i=j$, the real part of the Green's function diverges, corresponding to an unphysical self-interaction. We avoid this issue by simply setting $\Delta_{ii} = 0$, which may be viewed as a renormalization of the transition frequency. The imaginary part of the Green's function remains finite and describes spontaneous emission, $\Gamma_{ii} = \gamma_i$.

\section{Curved array mirrors}\label{ap:phasematching}

\begin{figure}[h!]
	\includegraphics[width=0.99\linewidth]{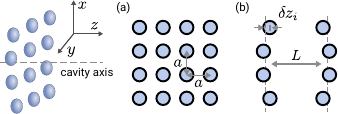}
	\caption{\label{curvsketch} Schematic depiction of the displacement of the atoms to realize curved array mirrors. (a) Projection onto the $z=0$ plane. The $x$ and $y$ positioning of the atoms remains a regular square lattice with spacing $a$. (b) $y=0$ plane cut. The atoms are displaced in the $z$ direction. The positions $z_i=\pm L/2+\delta z_i$ are determined by solving \eqref{eq:phasematching}.
 }
\end{figure}

For a cavity to support a Gaussian beam, the mirrors need to be curved to match the wavefront of the beam and refocus it at each reflection. To this end, we shift the position of the atoms along the $z$ direction corresponding to the axis of the optical cavity (see Fig.~\ref{curvsketch}). The position of the atoms in the ${x-y}$ plane remains unaltered. To determine the required displacement $\delta z$ of each mirror atom, we solve the phase-matching condition,
\begin{equation}\label{eq:phasematching}
    k_0z + k_0\frac{x^2+y^2}{2R(z)} - \psi(z) = k_0\frac{L}{2}\,,
\end{equation}
where $R(z) = (z^2+z_\mathrm{R}^2)/z$ is the radius of curvature of the wavefront at position $z$ and $\psi(z)=\arctan(z/z_\mathrm{R})$ is the Gouy phase of the fundamental Gaussian mode. Both quantities depend on the Rayleigh range $z_\mathrm{R} = k_0 w_0^2/2$, which is set by the wavenumber $k_0$ and the beam waist $w_0$. The left-hand-side of \eqref{eq:phasematching} corresponds to the phase of a Gaussian beam centered at the origin, which we match to the phase of a plane wave at the surface of a planar mirror of a cavity with length $L$ on the right-hand-side. Since \eqref{eq:phasematching} is a transcendental equation, we solve for the position $z_i$ corresponding to each lattice site $(x_i,y_i)$ numerically. Note that the distance between the arrays on the axis becomes slightly larger than $L$ to accommodate for the Gouy phase.

\section{Self-energy}\label{ap:selfenergy}
In this section, we justify the use of the spectral function \eqref{eq:response} to study the atom-array cavity and motivate the underlying physical intuition. To understand the interaction of the target atom with the atom-array cavity, we initialize the target atom in its excited state and let it evolve under the presence of the cavity. We consider a single target atom and do not include an external drive. Using the definitions introduced in \secref{sec:cavitymodes}, the Hamiltonian \eqref{eq:drivenHam} reads
\begin{align}
\begin{split}
    \hat{H} = (\omega_\mathrm{a}-\frac{i}{2}\gamma_\mathrm{a})\hat{\sigma}_\mathrm{a}^+\hat{\sigma}_\mathrm{a}^- + \sum_{i,j} \vec{H}^\mathrm{AA}_{ij}\,\hat{\sigma}_i^+\hat{\sigma}_j^- \\
    + \sum_i\left(\vec{H}^\mathrm{AT}_i\,\hat{\sigma}_i^+\hat{\sigma}_\mathrm{a}^- + \vec{H}^\mathrm{TA}_i\,\hat{\sigma}_\mathrm{a}^+\hat{\sigma}_i^-\right)\,.
\end{split}
\end{align}

We define the state $\ket{\psi(t)} = c_\mathrm{a}(t)\,\hat{\sigma}_\mathrm{a}^+\ket{0} + \sum_i c_i(t)\,\hat{\sigma}_i^+\ket{0}$, with $c_\mathrm{a}(0)=1$. 
By projecting the Schrödinger equation onto the different basis states, we obtain the equations of motion 
\begin{align}
    \dot{c}_\mathrm{a}(t) &= -i(\omega_\mathrm{a}-\frac{i}{2}\gamma_\mathrm{a})\,c_\mathrm{a}(t)
    - i\sum_i\vec{H}^\mathrm{TA}_i\,c_i(t)\,, \\
    \dot{c}_i(t) &= -i\sum_j \vec{H}^\mathrm{AA}_{ij}\,c_j(t) - i\vec{H}^\mathrm{AT}_i\,c_\mathrm{a}(t)\,.
\end{align}

This set of equations can be solved by performing a Laplace transform, which we define with an imaginary variable ${f(\omega)=\mathcal{L}\{f(t)\}(-i\omega)}$. The transformed equations of motion read
\begin{align}
    -i\omega&\,c_\mathrm{a}(\omega) - 1 =
     -i(\omega_\mathrm{a}-\frac{i}{2}\gamma_\mathrm{a})\,c_\mathrm{a}(\omega) -i\sum_i\vec{H}^\mathrm{TA}_i\,c_i(\omega) \label{eq:lapat}\\
    -i\omega&\,c_i(\omega) = - i\sum_j \vec{H}^\mathrm{AA}_{ij}\,c_j(\omega) - i\vec{H}^\mathrm{AT}_i\,c_\mathrm{a}(\omega)\,, \label{eq:lapcav}
\end{align}
Isolating $c_i(\omega)$ from \eqref{eq:lapcav} and substituting it into \eqref{eq:lapat}, yields
\begin{equation}\label{eq:resolvent}
    c_\mathrm{a}(\omega) = i\left[\omega-\omega_\mathrm{a}+\frac{i}{2}\gamma_\mathrm{a}-\Sigma_\mathrm{a}(\omega)\right]^{-1}\,,
\end{equation}
with the self-energy of the target atom $\Sigma_\mathrm{a}(\omega)$ as defined in \eqref{eq:selfE}. To obtain the time dynamics, we would then proceed to carry out the inverse Laplace transform, which coincides with the inverse Fourier transform in \eqref{eq:ca}.

If the cavity perturbs the target atom only weakly, such that the self-energy is flat across the range of frequencies probed by the atom, we can make the pole approximation, $\Sigma_\mathrm{a}(\omega)\approx\Sigma_\mathrm{a}(\omega_\mathrm{a})$. The solution to the time dynamics is that of an atom in free space with the resonance energy and decay rate modified by $\Sigma_\mathrm{a}(\omega_\mathrm{a})$.

For frequencies near strong poles of \eqref{eq:resolvent}, the solution is more involved. In this case, we can compare \eqref{eq:resolvent} to the equivalent solution using the effective Jaynes-Cummings Hamiltonian,
\begin{equation}\label{eq:resolventJC}
    c_\mathrm{a}^\mathrm{conv}(\omega) = i\left[\omega-\omega_\mathrm{a}+\frac{i}{2}\gamma_\mathrm{3D} - \frac{g^2}{\omega-\omega_\mathrm{c}+\frac{i}{2}\kappa}\right]^{-1},
\end{equation}
and observe that an isolated resonance in $\Sigma_\mathrm{a}(\omega)$ generates the same dynamics as conventional cavity QED. The imaginary part of the last term of \eqref{eq:resolventJC} is a Lorentzian with height $g^2/\kappa$, centered at the resonance energy of the cavity mode $\omega_\mathrm{c}$. Therefore, we can extract $g$, $\kappa$ and $\omega_\mathrm{c}$ from fitting a Lorentzian to $\mathrm{Im}\{\Sigma_\mathrm{a}(\omega\}$. As we discuss in \secref{sec:cavitymodes}, single eigenstates of the atom-array cavity are responsible for the observed cavity-like resonances. This allows us to also extract the cavity parameters from the Hamiltonian without the need for fitting.

Finally, by subtracting from $\Sigma_\mathrm{a}(\omega)$ the contribution of the cavity eigenstates responsible for a cavity resonance, which yields $\Sigma_{\mathrm{a,weak}}(\omega)$ that perturbs the target atom weakly, we can obtain the effective free-space decay of the target atom, $\gamma_\mathrm{3D} = \gamma_\mathrm{a} - 2\,\mathrm{Im}\{\Sigma_{\mathrm{a,weak}}(\omega_\mathrm{a})\}$.

\section{Raman scheme}\label{ap:raman}

In this section we derive the Hamiltonian governing the effective dynamics between the two long-lived states $\ket{\mathrm{g}_1},\ket{\mathrm{g}_2}$ of the Raman scheme presented in Fig.~\ref{figProt}(a). To this end, we adiabatically eliminate the excited state of the target atoms $\ket{\mathrm{e}}$ to second order in perturbation theory. Because we are interested in the regimes in which the atom-array cavity sustains a well-defined cavity mode, we approximate the array part of the Hamiltonian by a conventional cavity as described in \secref{sec:cavityQED}. When the free-space interaction between target atoms is much smaller than the atom--cavity coupling, as in the configurations discussed in \figref{figProt}, we can neglect the former and derive the effective dynamics for a single target atom. The extension to multiple atoms is straightforward.

The Hamiltonian including the Raman scheme described in \secref{sec:protocols}, in a frame in which $\ket{\mathrm{g}_2}$ rotates with the frequency of the Raman drive and $\hat{a}$ and $\ket{\mathrm{g}_1}$ rotate with $\omega_\mathrm{c}$, reads
\begin{equation}\label{eq:ramanevo}
\begin{split}
    \hat{H} &= -\frac{i}{2}\kappa\,\hat{a}^\dagger\hat{a} + (\omega_\mathrm{c} -\Delta_1+\Delta_2)\ket{\mathrm{g}_2}\!\bra{\mathrm{g}_2}\\
    &+ \omega_\mathrm{c}\ket{\mathrm{g}_1}\!\bra{\mathrm{g}_1} + (\omega_\mathrm{c} -\Delta_1-\frac{i}{2}\gamma_\mathrm{3D})\ket{\mathrm{e}}\!\bra{\mathrm{e}} \\
    &+ \frac{\Omega}{2}\left(\ket{\mathrm{g}_2}\!\bra{\mathrm{e}} + \mathrm{h.c.}\right) + g\left(\ket{\mathrm{g}_1}\!\bra{\mathrm{e}}\hat{a}^\dagger + \mathrm{h.c.}\right)\,.
\end{split}
\end{equation}
By shifting the energy of the atomic subspace by $(\Delta_1-\Delta_2)/2-\omega_\mathrm{c}$, the diagonal part of the Hamiltonian becomes
\begin{align}
\begin{split}
    \hat{H}_0 =& -\frac{i}{2}\kappa \,\hat{a}^\dagger\hat{a} + \frac{\Delta_2-\Delta_1}{2}\left(\ket{\mathrm{g}_2}\!\bra{\mathrm{g}_2} - \ket{\mathrm{g}_1}\!\bra{\mathrm{g}_1}\right) \\
    &- \left(\frac{\Delta_1+\Delta_2}{2}+\frac{i}{2}\gamma_\mathrm{3D}\right)\ket{\mathrm{e}}\!\bra{\mathrm{e}} .
\end{split}
\end{align}

For $|\Delta_1+\Delta_2|/2 \gg |\Delta_1-\Delta_2|/2, \Omega, g$, the state $\ket{\mathrm{e}}$ will only be weakly populated. We define the projector onto the subspace that includes $\ket{\mathrm{g}_1}$ and $\ket{\mathrm{g}_2}$,
\begin{equation}
    \hat{P} = (\ket{\mathrm{g}_2}\!\bra{\mathrm{g}_2} + \ket{\mathrm{g}_1}\!\bra{\mathrm{g}_1})\otimes \vec{I}_\mathrm{c}\,,
\end{equation}
and the complementary projection containing $\ket{\mathrm{e}}$,
\begin{equation}
    \hat{Q} = \vec{I}-\hat{P} = \ket{\mathrm{e}}\!\bra{\mathrm{e}}\otimes\vec{I}_\mathrm{c}\,,
\end{equation}
where $\vec{I}_\mathrm{c}$ is the identity matrix acting on the cavity subspace. Upon adiabatically eliminating $\ket{\mathrm{e}}$, the dynamics in the low-energy subspace are described to second order in perturbation theory by the Hamiltonian \cite{sternheim1972non,bravyi2011Schrieffer}
\begin{equation}
    \hat{H}_\mathrm{eff} = \hat{P}(\hat{H}_0+\hat{V})\hat{P}-\hat{P}\hat{V}(\hat{Q}\hat{H}_0\hat{Q})^{-1}\hat{V}\hat{P}\,,
\end{equation}
where
\begin{equation}
    \hat{V} = \frac{\Omega}{2}\ket{\mathrm{g}_2}\!\bra{\mathrm{e}} + g\ket{\mathrm{g}_1}\!\bra{\mathrm{e}}\hat{a}^\dagger + \mathrm{h.c.}
\end{equation}
is the part of the Hamiltonian that couples $\hat{P}$ and $\hat{Q}$.

The second-order term of the effective Hamiltonian reads
\begin{equation}
\begin{split}
    \bigg(\frac{\Omega}{2}\ket{\mathrm{g}_2}+g\ket{\mathrm{g}_1}\hat{a}^\dagger\bigg)&\\
    \times\bigg(\frac{1}{2}(\Delta_1+\Delta_2)+&\frac{i}{2}\gamma_\mathrm{3D}+\frac{i}{2}\kappa\hat{a}^\dagger\hat{a}\bigg)^{-1}\\
    &\times \bigg(\frac{\Omega}{2}\bra{\mathrm{g}_2}+g\bra{\mathrm{g}_1}\hat{a}\bigg)\,.
\end{split}
\end{equation}
In the main text, we only consider the single-excitation subspace $\{\ket{\mathrm{g}_2}\ket{0}_\mathrm{c},\ket{\mathrm{g}_1}\ket{1}_\mathrm{c}\}$, with $\ket{1}_\mathrm{c} = \hat{a}^\dagger\ket{0}_\mathrm{c}$, for which the $i\kappa\hat{a}^\dagger\hat{a}/2$ term in the denominator equals zero. This term can also be safely neglected for a small number of photons in the cavity if $\kappa\ll|\Delta_1+\Delta_2|/2,\gamma_\mathrm{3D}$. We will therefore neglect this term in what follows.

The resulting effective Hamiltonian after adding a global energy shift $(\Delta_2-\Delta_1)/2$ to the atomic states reads
\begin{equation}\label{eq:ramanfull}
	\begin{split}
	\hat{H}_\mathrm{eff} &= \left(-\frac{i}{2}\kappa  + \frac{2g^2}{(\Delta_1+\Delta_2)+i\gamma_\mathrm{3D}}\ket{\mathrm{g}_1}\!\bra{\mathrm{g}_1}\right)\hat{a}^\dagger\hat{a} \\
	& + \left(\Delta_2-\Delta_1 + \frac{\Omega^2/2}{(\Delta_1+\Delta_2)+i\gamma_\mathrm{3D}}\right)\ket{\mathrm{g}_2}\!\bra{\mathrm{g}_2} \\
    & - \frac{g\,\Omega}{(\Delta_1+\Delta_2)+i\gamma_\mathrm{3D}}\left(\ket{\mathrm{g}_2}\!\bra{\mathrm{g}_1}\hat{a} + \mathrm{h.c.}\right) \,.
	\end{split}
\end{equation}
The two-photon transition connecting $\ket{\mathrm{g}_1} \leftrightarrow \ket{\mathrm{g}_2}$ is resonant for $\Delta_1\approx\Delta_2$ up to small Stark shift corrections. In the regime $|\Delta_1|\gg\gamma_\mathrm{3D}$, the dynamics of the effective target atom are well approximated by a two-level atom with the reduced free-space decay rate and coupling strength with the cavity mode defined in \secref{sec:protocols}. Note that for smaller $|\Delta_1|\sim\gamma_\mathrm{3D}$, the first and third line of \eqref{eq:ramanfull} contain additional dissipative (anti-Hermitian) terms. By requiring that the corresponding rates are much smaller than the cavity decay rates, we can obtain a more precise condition for $\Delta_1$ in relation to $\gamma_\mathrm{3D}$. In particular we need $g^2 \gamma_\mathrm{3D} / \Delta_1^2 \ll \kappa$ and $g \Omega \gamma_\mathrm{3D} / \Delta_1^2 \ll \kappa$. Using $\Omega/\Delta_1=\sqrt{\kappa/\gamma_\mathrm{3D}}$, both conditions are satisfied when $|\Delta_1|\gg \sqrt{C}\gamma_{3D}$.

\section{Transmission and reflection}\label{ap:transmission}
To compute the transmission and reflection coefficient of an array, we drive the array with a field $\vec{E}_0^+$ and reconstruct the total field using the input-output formula in \eqref{eq:IO}. When discussing transmission and reflection of the mirrors of a cavity, it is important that the reflected light beam retains its spatial profile. For experimental realizations, it is also desirable that the transmitted beam is in some target collimated mode. Therefore, to distinguish between transmission (reflection) and scattering into arbitrary directions, we project the total field onto a detection mode $\vec{E}^+_\mathrm{det}(\vec{r})$ ($\vec{E}^-_\mathrm{det}(\vec{r})$). Following a similar approach to previous works~\cite{manzoni2018optimization,pedersen2023quantum} and working within the paraxial approximation, we obtain the transmission and reflection coefficients
\begin{align}
\begin{split}
    t =& \int \mathrm{d}x\,\mathrm{d}y\, \vec{E}_\mathrm{det}^-(\vec{r})\cdot\vec{E}^+_0(\vec{r}) \\
    &+ \frac{i}{2k_0}\frac{\omega_0^2}{\epsilon_0 c^2} \sum_i \vec{E}^-_\mathrm{det}(\vec{r}_i)\cdot\vec{d}_i \braket{\hat{\sigma}_i^-}
\end{split}\\
    r =& \,\frac{i}{2k_0}\frac{\omega_0^2}{\epsilon_0 c^2} \sum_i \vec{E}^+_\mathrm{det}(\vec{r}_i)\cdot\vec{d}_i \braket{\hat{\sigma}_i^-}
\end{align}
where the input and detection fields are normalized such that $\int \mathrm{d}x\,\mathrm{d}y \,\vec{E}^-_\alpha(\vec{r})\cdot\vec{E}^+_\alpha(\vec{r})=1$. For both input field and detection mode, we use the same Gaussian beam 
\begin{equation}
\begin{split}
    \vec{E}_\mathrm{G}^+(\vec{r}) =& \,\vec{e}_x\,\sqrt{\frac{2}{\pi}}\frac{1}{w(z)}\exp\left(-\frac{x^2+y^2}{w(z)^2}\right)\\
    &\times\exp\left[-i\left(k_0z+k_0\frac{x^2+y^2}{2R(z)}-\psi(z)\right)\right]\,,
\end{split}
\end{equation}
with a linear polarization aligned with the atom dipoles. The beam waist $w_0$ is the same as the waist chosen to determine the curvature of the array mirrors. The quantities $R(z)$ and $\psi(z)$ are as defined in \secref{ap:phasematching} and $w(z)=w_0\sqrt{1+(2z/k_0w_0^2)^2}$.

For $\braket{\hat{\sigma}_i^-}$, we use the steady state solution of the internal dynamics, $\braket{\dot{\hat{\sigma}}_i^-} = 0$. Assuming a weak drive, such that saturation of the atoms is negligible, the Heisenberg equations of motion from the Hamiltonian \eqref{eq:drivenHam} read
\begin{equation}
    \braket{\dot{\bm{\sigma}}^-(t)} = -i\,\vec{h}\cdot\bm{\sigma}^-(t) + i\,\bm{\Omega}\,,
\end{equation}
where we have written the Hamiltonian in matrix form, $\hat{H} = \bm{\sigma}^+\cdot\vec{h}\cdot\bm{\sigma}^- - (\bm{\Omega}\cdot\bm{\sigma}^++\mathrm{h.c.})$. The steady state solution is, thus,
\begin{equation}
    \bm{\sigma}^- = \vec{h}^{-1}\cdot\bm{\Omega}\,.
\end{equation}
We recall that $\Omega_i = \vec{d}_i^* \cdot \vec{E}_0^+(\vec{r}_i)$.

In the text, we also discuss the field associated with an eigenstate $\xi$ of the system. In this case, we reconstruct the field emitted by $\braket{\hat{\sigma}_\xi}$ using \eqref{eq:IO}, and evaluate the field intensity $\braket{\vec{E}^-(\vec{r})\vec{E}^+(\vec{r})}$, using that in the linear, unsaturated, regime $\braket{\vec{E}^-(\vec{r})\vec{E}^+(\vec{r})} \approx \braket{\vec{E}^-(\vec{r})}\braket{\vec{E}^+(\vec{r})}$.

\section{Effective Hamiltonian in the fast-motion regime}\label{ap:fastmotion}
In the fast-motion regime, the effective Hamiltonian should be averaged over the fluctuations of the positions of the atoms. There are two distinct quantities in \eqref{eq:drivenHam} that depend on the position of the atoms: the interaction coefficients $\Delta_{ij} - i \Gamma_{ij} / 2$ and the Rabi frequencies $\Omega_i$. For both quantities, the average can be evaluated analytically in the limit that the fluctuations are small compared to the wavelength $\lambda_0$ and the lattice spacing $a$.

For the interaction coefficients, we have to compute the average of the Green's function $\vec{G}(\vec{r}_i + \vec{r}, \vec{r}_j + \vec{r'}; \omega_0)$, where $\vec{r}$ and $\vec{r'}$ are drawn independently from the probability distribution $p(\vec{r}) = e^{-r^2 / 2 \sigma^2} / (2 \pi \sigma^2)^{3/2}$. To simplify notation, we introduce the shorthand $\vec{G}(\vec{r}_i - \vec{r}_j) = \vec{G}(\vec{r}_i, \vec{r}_j; \omega_0)$. We obtain the averaged Green's function
\begin{align}
    \bar{\vec{G}}(\vec{r}_i - \vec{r_j}) &= \int \di^3 \vec{r} \int \di^3 \vec{r'} p(\vec{r}) p(\vec{r'}) \vec{G}(\vec{r}_i + \vec{r} - \vec{r}_j - \vec{r'}) \nonumber\\
    &= \frac{1}{(4 \pi \sigma^2)^{3/2}} \int \di^3 \vec{r} \, \vec{G}(\vec{r}_i - \vec{r}_j + \vec{r}) e^{-r^2 / 4 \sigma^2} \nonumber\\
    &\approx \vec{G}(\vec{r}_i - \vec{r}_j) + \sigma^2 \left. \nabla^2 \vec{G}(\vec{r}) \right|_{\vec{r} = \vec{r}_i - \vec{r}_j},
\end{align}
where the last line follows from expanding $\vec{G}(\vec{r}_i - \vec{r}_j + \vec{r})$ to second order in $\vec{r}$, corresponding to a saddle-point approximation. To evaluate the Laplacian, we make use of the definition of the Green's function,
\begin{equation}
    \nabla \times \nabla \times \vec{G}(\vec{r}) - k_0^2 \vec{G}(\vec{r}) = \delta(\vec{r}) \vec{I} .
\end{equation}
At $\vec{r} \neq 0$, the Green's function further satisfies $\nabla \cdot \vec{G}(\vec{r}) = 0$ because $\vec{G}(\vec{r})$ is an electric field, which is divergence free in the absence of sources. Using $\nabla \times \nabla \times \vec{G} = \nabla (\nabla \cdot \vec{G}) - \nabla^2 \vec{G}$, the Green's function therefore satisfies
\begin{equation}
    \left( \nabla^2 + k_0^2 \right) \vec{G}(\vec{r}) = 0, \qquad \text{if } \vec{r} \neq 0 .
\end{equation}
Combining everything yields
\begin{align}
    \bar{\vec{G}}(\vec{r}_i - \vec{r_j}) &\approx (1 - k_0^2 \sigma^2) \vec{G}(\vec{r}_i - \vec{r_j}) \nonumber\\
    &\approx e^{-k_0^2 \sigma^2} \vec{G}(\vec{r}_i - \vec{r_j}) .
\end{align}
This justifies the rescaling factor of the last term in \eqref{eq:fastmotion}.

A very similar argument applies to the Rabi frequency. Within the saddle-point approximation, the averaged Rabi frequency is given by
\begin{align}
    \bar{\Omega}_i &= \int \di^3 \vec{r} \, p(\vec{r}) \, \vec{d}_i^* \cdot \vec{E}(\vec{r}_i + \vec{r})\\
    &\approx \Omega_i + \frac{1}{2} \sigma^2 \left. \vec{d}_i^* \cdot \nabla^2 \vec{E}(\vec{r}) \right|_{\vec{r} = \vec{r}_i} .
\end{align}
We again use $(\nabla^2 + k_0^2) \vec{E}(\vec{r}) = 0$ to obtain
\begin{align}
    \bar{\Omega}_i \approx e^{-k_0^2 \sigma^2 / 2} \, \Omega_i .
\end{align}
This is the second term in \eqref{eq:fastmotion}.

\section{Extended description of \figref{figAtomConv}}\label{ap:fig10}
The ``ideal'' points are computed for cavities with curved arrays. We consider lattice spacings $a=0.47\lambda_0$, $a=0.57\lambda_0$, and $a=0.68\lambda_0$ and sizes $N\times N=60\times60$, $N\times N=50\times50$, and $N\times N=40\times40$, respectively, for all combinations of $w_0/\lambda_0 \in \{2.5, 3.8, 5.0 \}$, and $L/\lambda_0 \in \{5.5, 20.5, 50.5 \}$. We also computed the results for $L=1.5\lambda_0$. They are very similar to the ones for $L=5.5\lambda_0$, except when $a=0.68\lambda_0$, for which the target atom experiences a strong near-field interaction with the arrays, departing from the optical cavity regime (see \figref{figCoupStrength}, right panel).

Apart from a few exceptions, the ``ideal'' points are consistently above the theory line by a factor of roughly $3/2$. We attribute this shift to differences between the mode profile and the Gaussian beam used to probe the mirror's reflectivity [see \figref{figMaxTransmission}(c)], leading to a systematic underestimate of $\mathcal{R}$. The outliers below the theoretical prediction are likely due to the coupling between different cavity modes. The ``ideal'' points furthest below the dashed line for the three values of $a$ correspond to cavities with $w_0=2.5\lambda_0$ and $L=50.5\lambda_0$. These points have the largest value of the ratio $L/R$, where $R$ is the radius of curvature of the mirror. A large value of $L/R$ leads to a small separation between $\mathrm{TEM}_{mn}$ modes, as discussed in \secref{sec:gamma3D}. The small splitting results in increased mixing of the transverse modes, which is not accounted for in the computation of the reflection coefficient of a single array mirror.

The ``transmittive'' points include the data of the cavities shown in \figref{figMaxTransmission}(b) in the range $L / \lambda_0 \in[5.45,5.50) $. In this case, the modes have Gaussian profiles, so the agreement with the theory is excellent. The increased $\kappa_\mathrm{out}$ of these cavities comes naturally at the expense of smaller $C$. The ``motion'' points include the data shown in \figref{figMotion} in the frozen motion regime, for which $\kappa_\mathrm{out}$ is very small. The points that are colored as half ``motion'' and half ``transmittive'' correspond to the points shown in \figref{figIntro}(d). The ``Stark shift'' points correspond to the data shown in \figref{figStark}(c--e). The top half of the points are for $L=1.5\lambda_0$, and the bottom half for $L=5.5\lambda_0$. We show points in the range starting with $w_0\approx6\lambda_0$ down to $w_0\approx2\lambda_0$. We do not show the points with larger beam waists, as they have strong diffraction losses. The profile of the cavity mode with Stark-shifted mirrors differs the most from a Gaussian mode, which explains the larger displacement from the dashed line.

\bibliography{bibliography}

\end{document}